\documentclass[graybox, nosecnum]{svmult}

\usepackage{mathptmx}       
\usepackage{helvet}         
\usepackage{courier}        
\usepackage{type1cm}        
\usepackage{makeidx}         
\usepackage{graphicx}        
\usepackage{multicol}        
\usepackage[bottom]{footmisc}
\usepackage{hyperref}        
\usepackage{soul}            
\hypersetup{colorlinks=true,urlcolor=blue}
\usepackage[square,sort,comma,numbers]{natbib}

\usepackage{sidecap}
\usepackage{tabularx}
\makeindex             

\usepackage{a4wide,graphicx,subfigure,float}
\usepackage{empheq}

\newcommand{\Eqref}[1]{Eq.~(\ref{#1})}
\newcommand{\Eqsref}[2]{Eqs.~(\ref{#1}) and (\ref{#2})}

\newcommand{\figref}[1]{Fig.~(\ref{#1})}

\newcommand{\beq}[1]{\begin{equation}\label{#1}}
\newcommand{\eeq}{\end{equation}}

\title*{Transition-Edge Sensors  for cryogenic X-ray imaging
spectrometers}

\author{Luciano Gottardi\thanks{corresponding author} and  Stephen Smith}
 \institute{Luciano Gottardi \at NWO-I/SRON Netherlands Institute for Space Research, Niels Bohrweg 4, 2333CA Leiden, The Netherlands. \email{l.gottardi@sron.nl} 
      \and 
      Stephen Smith \at NASA Goddard Space Flight Center, 8800 Greenbelt Road, Greenbelt, MD 20771, USA \email{stephen.j.smith@nasa.gov}
         }
\begin{document}
\maketitle
\abstract{Large arrays of superconducting transition-edge sensor (TES)  microcalorimeters are becoming the key technology for future space-based X-ray observatories and  ground-based experiments in the fields of astrophysics, laboratory astrophysics, plasma physics, particle physics and material analysis. Thanks to their sharp superconducting-to-normal transition, TESs can achieve very high sensitivity in detecting small temperature changes at very low temperature. TES based X-ray detectors are non-dispersive spectrometers bringing together high resolving power, imaging capability and high-quantum efficiency simultaneously. In this chapter, we highlight the basic principles behind the
operation and design of TESs, and their fundamental noise limits. We will further elaborate on  the key fundamental physics processes that guide the design and optimization
of the detector. We will then describe pulse-processing and important calibration
considerations for space flight instruments, before introducing novel multi-pixel TES
designs and discussing applications in future X-ray space missions over the coming decades.}
\section{Keywords} 
\begin{keywords}Imaging spectroscopy, transition-edge sensor, microcalorimeter array, superconductivity, weak-link, X-ray astrophysics, laboratory astrophysics.
\end{keywords}
\tableofcontents{}
\section{Introduction}

X-ray spectroscopy provides an excellent diagnostic probe of the energetics and physical conditions in nearly all classes of cosmic objects throughout the observable universe. X-rays are emitted by various high-energy processes. By observing their spectra we can obtain information about the temperature, electron density and ionic composition of hot  plasma's, and answer many questions across astrophysics (from understanding turbulence in galaxy clusters to the accretion processes in binary systems or active galactic nuclei.)

Microcalorimeters are non-dispersive thermal detectors that will provide the next major step in imaging spectroscopy capabilities compared to gas proportional counters or charge coupled devices (CCD) which have been extensively used in X-ray space instrumentation over the past several decades. Microcalorimeters will full-fill the needs of X-ray astrophysics in the 21st century, combining eV level energy resolution in the soft X-ray energy range in large format imaging arrays with potentially 1000's of pixels. With resolving powers $>$ 1000, microcalorimeters are competitive with dispersive grating spectrometers, but with the advantage of high collection efficiency. This enables efficient observations of point sources and extended sources with the same instrument. The sensor technology used in microcalorimeters can come in different forms - silicon thermistors, transition-edge sensors (TESs) and magnetic microcalorimeters (MMCs) - but the basic principle is the same. A microcalorimeter measures the temperature rise resulting from the energy deposited by a single X-ray photon. The sensor transduces the change in temperature to an electrical signal (either through a resistance change for thermistors and TESs or a change in magnetism for MMCs), from which the photon energy can then be determined. In order to achieve resolving powers of $\sim$1000's at keV energies, extremely low detector noise is required, only achievable when operating at mK temperatures. The potential power of microcalorimeters to revolutionize X-ray astrophysics has already been demonstrated by the observation of the Perseus galaxy cluster by the Hitomi satellite's Soft X-ray Imaging Spectrometer (SXS) \cite{Hitomi2016}. 

In this chapter, we describe microcalorimeters based around highly sensitive transition-edge-sensors (TESs). TESs are next generation microcalorimeters based on a thin superconducting film that is electrically biased in the narrow region between superconducting (zero-resistance) state and the normal resistive state. The transition region is typically only a few mK in width and as such, TES detectors are extremely sensitive to temperature changes making them ideal detectors for high resolution X-ray spectroscopy. We start by outlining the basic principles behind the operation and design of TESs, and their fundamental noise limits. We will elaborate on  the key fundamental physics effects that guide the design and optimization of the TESs. We will then describe important calibration considerations for space flight instruments.  We will continue by  introducing  novel multi-pixel TES designs and conclude the chapter by presenting the applications in future X-ray space missions over the coming decades.

\vspace{0.3cm}

\section{Theoretical and Experimental Background}
\subsection{Basic Principles}\label{Basic_principles}
\begin{figure}
\center
\includegraphics[width=5.5cm,angle=270]{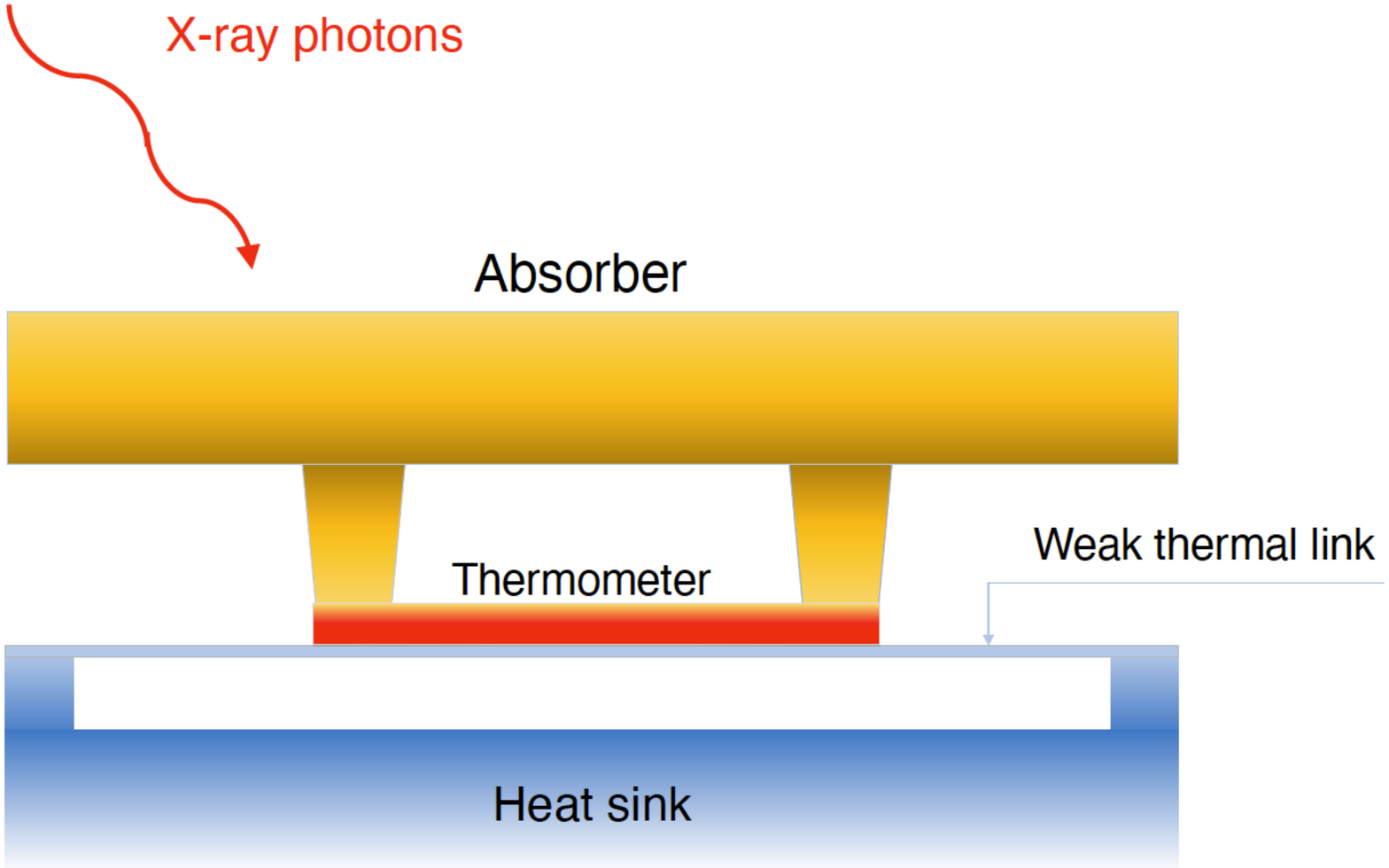}
\caption{\label{fig:tescal} Schematic of a TES-based X-ray microcalorimeter. The TES is a very sensitive thermometer that detects the temperature rise from the energy deposited in the absorber by an X-ray photon.}
\end{figure}

A TES-based X-ray detector is a thermal equilibrium calorimeter. It consists of three fundamental components: an absorber with heat capacity $C$, a thermometer (the TES), and a weak thermal link, with a conductivity $G_{bath}$, to the heat sink at temperature $T_{bath}$, which is below the operating temperature of the device. Details on the design of each of these components is described in Section \ref{Detector_design}. A simplified scheme of a TES calorimeter is shown in \figref{fig:tescal}. 

TESs are based on superconducting thin films, voltage biased in the narrow transition region between the onset of resistance and the fully normal state. A single photon deposits its energy $E$ into the absorber, which  converts it into heat. The temperature rise, proportional to the energy, causes a change in resistance of the TES. The resistance change is determined by monitoring the current through the TES using a superconducting quantum interference device (SQUID) ammeter. Within this general description, there is room for countless variations. Important device parameters such as the noise, $G_{bath}$ and $C$ are strongly dependent on the operating temperature of the device. Thus, the transition temperature of the TES must be chosen to achieve the desired energy resolution and response time, whilst being compatible with the instrument cryogenic architecture. For typical astrophysics applications a transition temperature of below 100 mK is ideal.

Thermal equilibrium detectors can achieve excellent energy resolution. A~fundamental limit for the minimum energy resolution $\Delta E$ offered by a calorimeter is given by the random exchange of energy between the detector and the thermal bath~\cite{McCammon2005}. This thermodynamic limit is given by $<\Delta E_{TD}^2>=k_BT^2C$. It depends quadratically on the temperature $T$ of the calorimeter, linearly on the detector heat capacity $C$, and it is independent on the thermal conductance $G_{bath}$ of the thermal link.

The ultimate sensitivity of a TES microcalorimeter depends on the shape of the TES superconducting-to-normal transition and the intrinsic noise sources of the detector and the read-out circuit. The resistance, $R(T,I)$, of a TES is generally a function of both the temperature of the device, $T$, and the current flowing through it, $I$. For small changes about the equilibrium bias conditions ($R_0,T_0,I_0$) the device resistance can be expressed as
\begin{equation}\label{Ralphabeta} 
R(T,I) \simeq R_0+\alpha\frac{R_0}{T_0}\delta T +\beta\frac{R_0}{I_0}\delta I.
\end{equation}
The two dimensionless parameters  $\alpha=\partial\log R/\partial\log T$ and $\beta=\partial\log R/\partial\log I$, calculated  at a constant current and temperature respectively, are conveniently used to parameterize the current and temperature sensitivity of the resistive transition at the operating point~\cite{Lind2004}.

The energy resolution
of an ideal TES calorimeter, limited only by the fundamental thermal fluctuations
due to the exchange of energy between the calorimeter and the thermal bath,
is given by \cite{McCammon2005}
\begin{equation}\label{eq:dE}
\Delta E \propto \sqrt{4k_BT^2_0\frac{C}{\alpha}},
\end{equation}
where the logarithmic derivative of resistance $\alpha$, introduced above, describes the steepness of the superconducting transition. By~developing TESs with a large temperature sensitivity $\alpha$, the~photon energy can be measured with a much higher resolution than the magnitude set by thermodynamic fluctuations. \Eqref{eq:dE} tells us that  low heat capacity devices,  operating at very low temperature $T_0$, could achieve very high energy resolution. However, the value of $C$,$\alpha$ and $T_0$ are constrained by the maximum energy to be detected. For high energetic photons, the temperature excursion could be so high to drive the TES in the normal state and to saturate the detector. The saturation energy $E_{sat}$ is proportional to $E_{sat}\propto C/\alpha$, which effectively means that the theoretical energy resolution $\Delta E \propto \sqrt{E_{sat}}$.

A detail analysis of the noise, the electrothermal response and the ultimate sensitivity of a TES microcalorimeter will be presented in the following sections.

\subsection{TES Electrical and Thermal Response}
A TES detector can be electrically voltage biased in its transition region using either an AC or DC voltage biasing scheme. Although the design of the electrical bias circuit is naturally different for the two approaches, the resulting electrothermal behaviour of the TES is generally equivalent. Whether a TES is AC or DC biased can, however, affect the transition shape and noise properties of the device and may lead to different optimal pixel designs \cite{GottNaga2021}.

\begin{figure}
\center
\includegraphics[width=8.5cm]{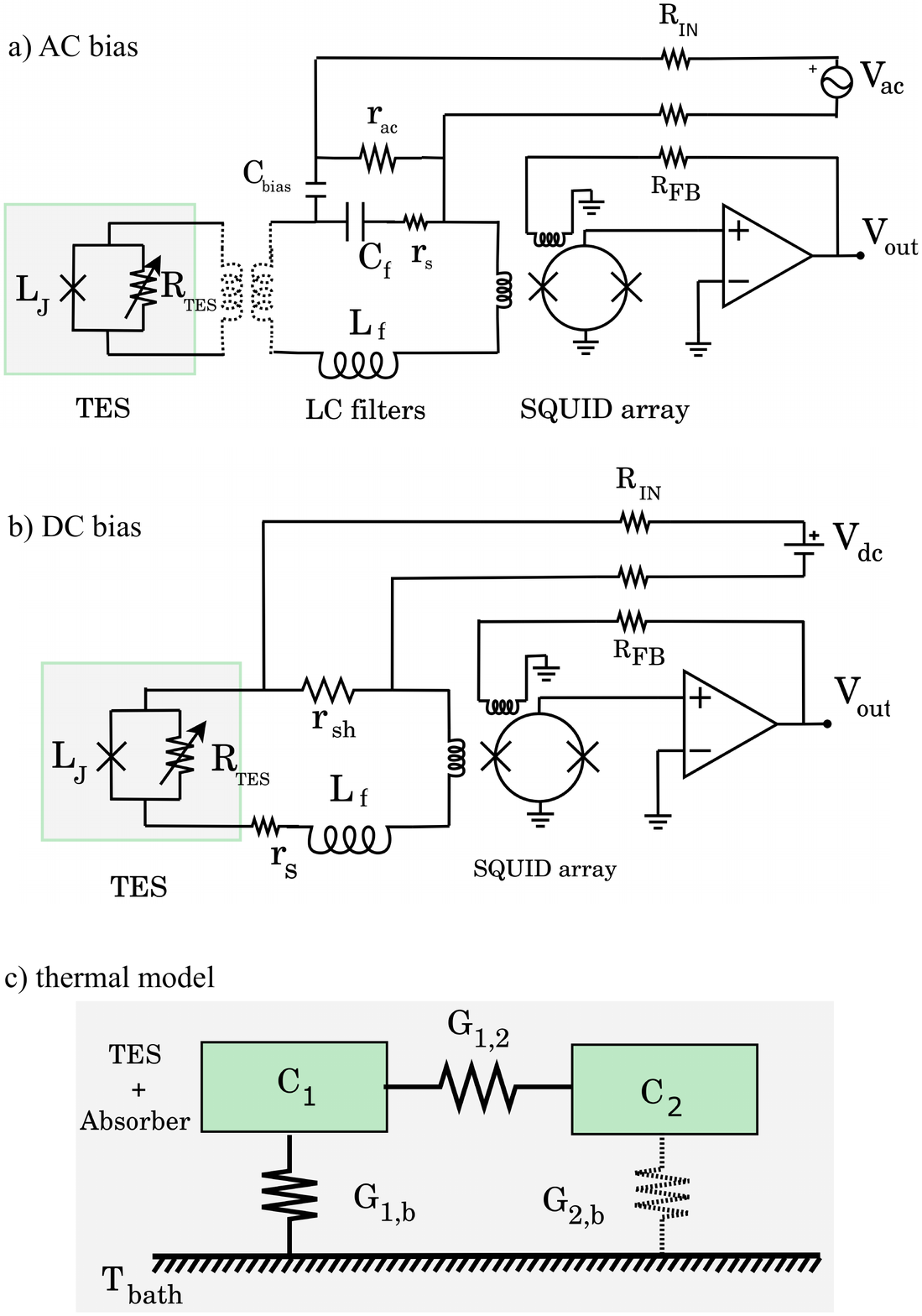}
\caption{\label{fig:ETscheme} a) AC bias electrical circuit, b) DC bias electrical circuit, c) Example of a two-body thermal model with two heat capacitance $C_1$  and $C_2$, linked by a thermal conductance $G_{12}$, and with parallel connection to the heat sink, $G_{1,b}$ and $G_{2,b}$, defining the total thermal conductance to the bath, $G_{bath}$ .}
\end{figure} 
A~schematic diagram of the AC and DC bias read-out circuit  and the detector thermal model are  shown in  \figref{fig:ETscheme}. 
The fundamental difference between the two bias configurations is that, in the AC bias case, the TES is brought into transition by the rms power generated by an alternating voltage  at few MHz frequency, while in the DC bias case a constant voltage is applied.  

Under AC bias the~TES is placed in series with a high-$Q$ superconducting  $LC$ filter~\cite{Bruijn18,Bruijn14} and the input coil of the SQUID current amplifier~\cite{Kiviranta02}. An~optional superconducting transformer, shown in the picture in dashed lines, could be used to optimize the impedance matching between detector and amplifier. The~AC voltage bias, tuned at the $LC$ filter resonant frequency $f_{bias}=f_{LC}=1/(2\pi\sqrt{LC})$, is provided via a capacitive divider in parallel with an AC shunt resistor. From~a simple Thevenin equivalent circuit analysis, the~bias network is equivalent to a shunt resistance  $r_{sh}\simeq r_{ac}C_b/(C_F+C_b)+r_s$, in~series with the $LC$ resonator, where $r_s$ accounts for  the intrinsic loss in the $LC$-filters~\cite{Gottardi19}.
The DC bias circuit is shown in \figref{fig:ETscheme}b. where $L_f$ is the Nyquist inductor added to limit the read-out bandwidth, $r_{sh}$ is a low ohmic shunt resistors needed to provide a stiff dc voltage bias to the TES, and $r_s$ indicates any parasitic resistance in the bias circuit. Both for the AC and DC read-out the following should hold: $r_s\ll r_{sh} \ll R_N$.
The standard electrothermal linear model that describes the noise and the small signal response to photons of a DC biased TES calorimeter has been extensively presented and analysed in the literature \cite{IrwinHilton,Lind2004}. The formalism for the AC biased case was first reported by van der Kuur, J. et al.  in \cite{JvdKuur11}. In this model, the TES resistance is linearly approximated as in \Eqref{Ralphabeta} with  no assumptions made on the physical  phenomena occurring at the superconducting  transition. This simplified model has been successfully used to understand the behaviour of TES-based detectors in many  works that followed.
As observed in many experiments~\cite{Takei08,Kinn12,Maasilta12,Goldie_2009,Wake2019}, due to the presence of parasitic heat capacitance and a limited thermal conductance in the TES-absorber structures, a~single-body thermal model for  a TES calorimeter is not always sufficient to explain the detector thermal response.  It can be shown that a two-body model, as~drawn  in \mbox{\figref{fig:ETscheme}c} with two heat capacitance $C_1$  and $C_2$ linked by a thermal conductance $G_{12}$, is sufficiently general to account for parasitic thermal effects in TES-based detectors. 
The heat capacity $C_1$  is typically the sum of the TES bilayer and the absorber heat capacity, since the two structures are generally thermally well coupled, while $C_2$ is a potential decoupled heat capacitance which could have different physical sources, like a fraction of the TES bilayer, the~supporting $Si_xN_y$ membrane, or the leads~\cite{Takei08, Goldie_2009,Wake2019, Lindeman_2011}.
A~more detailed analysis for even more complex thermal structures can be found for example \mbox{in~\cite{Kinn12,Maasilta12}}. One has to keep in mind, however, that the model could be  unconstrained when too many thermal bodies are added into the system of equations. As~a consequence, the~physical interpretation of the results becomes~impossible. 

The detector response is regulated by the Joule power provided by the bias circuit, $P_J=V_{TES}^2/R_{TES}$ and the power $P_{bath}$ flowing to the thermal bath, given by 
 \beq{eq:Pbath}
 P_{bath}=k(T^n-T^n_{bath}),
 \eeq
where $k=G_{bath}/n(T^{n-1})$, with~$G_{bath}$ the differential thermal conductance to the thermal bath, $n$ the thermal conductance
exponent, and $T_{bath}$ the bath temperature~\cite{IrwinHilton}. The parameters $n$ and $K$ are material and geometry dependent parameters that depend on the design of the physical link to the heat bath (see Section \ref{Thermal_isolation}). The system is characterized by two fundamental time constants, $\tau=C/G_{bath}$ and $\tau_{el}=\frac{L}{r_{sh}+R_0(1+\beta)}$, defining respectively the thermal and electrical bandwidth. 

The Langevin electrothermal equations  for a TES calorimeter, generalized for the DC and  AC biased case have been extensively studied in the literature \cite{IrwinHilton,Lindeman_2011,Ullom_2015,Wake2019,GottNaga2021}. The system of coupled non-linear differential electrothermal equations  is typically solved in the small signal regime and in the linear approximation. In~this case, the detector physics is generally ignored  and $R(T,I)$ is parametrized as in \Eqref{Ralphabeta} \cite{IrwinHilton,Lind2004,Lindeman_2011,JvdKuur11,TaralliAIP2019}. For this system of coupled linear differential equations there exist analytical solutions, which are either over-,under-, or critically damped. The stability criteria is discussed in the following section.

As it will be shown  in Section \ref{sec:TESPhysics}, many of the physical phenomena observed at the TES transition can be explained in the framework of the weak-superconductivity. The detector physics can then be formally included in the electrothermal equations as shown in \cite{GottNaga2021}. They have been fully solved numerically, for~an AC biased pixel,  in~the simplified single-thermal body case~\cite{Kirsch2020}.  This numerical time-domain model can  be used to simulate the TES response to the incoming photons in the non-linear and large signal regime, as it is done for example in the end-to-end simulator under development for the X-IFU  on Athena~\cite{Wilms2016,Lorenz2020}.

\subsection{Negative Electrothermal Feedback}
\label{sec:ETF}
The thermal and electrical circuits of a TES are coupled via the cross-terms in the thermal-electrical differential equations \cite{GottNaga2021}. A temperature change in the TES leads to an electrical current signal as a result of a change in the TES resistance. Under voltage-bias conditions ($r_{sh}\ll R(T_0,I_0)$), the electrical current signal change is restored by a change of the Joule power, $P_J=V_{TES}^2/R(T_0,I_0)$, which decreases with increasing temperature. This process is analogous to the electrical feedback in a transistor circuit. It involves both the electrical and the thermal circuits simultaneously, and it has been named, for this reason, electrothermal feedback (ETF). In the voltage bias case, the ETF is negative and the TES is stable against thermal runaway. 
The zero frequency ETF loop gain $\mathcal{L}_0$ depends on the detector intrinsic quiescent parameters such as $\alpha$, $G_{bath}$, $P_0=I^2_0R_0$,  $T_0$ and the bias circuit shunt resistor $r_{sh}$, and it is defined as 

\beq{Lgain}
\mathcal{L}_0=\alpha\frac{1-r_{sh}/R(T_0,I_0)}{1+r_{sh}/R(T_0,I_0)+\beta} \frac{I^2_0R_0}{T_0 G_{bath}}. 
\eeq
For a constant current, $\beta=0$, and in the limit of $R(T,I)\gg r_{sh}$, the loop gain $\mathcal{L}_0$ reduces to the standard loop gain  $\mathcal{L}_I=\alpha\frac{I^2_0R_0}{T_0 G_{bath}}$ found in the literature \cite{IrwinHilton}.

As it is the case for transistors, there are significant advantages in operating a TES in the negative EFT mode, such as reduced
sensitivity to TES parameter variation, faster response time, increased linearity and dynamic range, self-biasing, and self-calibration. \cite{Irwin1995,IrwinHilton}. 
As any feedback scheme, ETF can become unstable depending on the detector and the bias circuit parameters. The stability criteria for a TES micro-calorimeter have been extensively discussed by Irwin \cite{Irwin1995,IrwinHilton}. For a single thermal body, the stability condition for  the solution of the system  is met for positive values of the real part of the eigenvalues $1/\tau_+$ and $1/\tau_-$ derived in \cite{IrwinHilton}, where $\tau_+$ and $\tau_-$ are defined as the the rise-time and fall-time of a pulse generated by a X-ray  photon absorbed by the detector.

In the limit of a small electrical inductance $L$, the rise-time $\tau_+$ reduces to the electrical circuit time constant $\tau_{el}$, while the  effective thermal time constant of the TES becomes
\beq{eq:taueff}
\tau_{eff}=\tau_-|_{L=0}=\frac{C}{G_{bath}}\frac{1+\beta+r_{sh}/R(T_0,I_0)}{1+\beta+r_{sh}/R(T_0,I_0)+(1-r_{sh}/R(T_0,I_0))\mathcal{L}_I}.
\eeq
In the stiff voltage bias limit ($r_{sh}<<R_0$), we get
\beq{eq:teff_aprox}
\tau_{eff}=\frac{C}{G}\frac{1+\beta}{1+\beta+\mathcal{L}_I}.
\eeq 
The equations above show that the relaxation time to the steady state after a photon absorption is reduced by the ETF loop gain $\mathcal{L}_I$. Furthermore,  the thermal time constant of the detector is directly proportional to
the $\alpha$ and $\beta$ parameters, and therefore depends significantly on the local properties of the transition.

The condition on the time constants $\tau_{\pm}$ being real for a stable, critically ($\tau_+=\tau_-$) or overdamped ($\tau_+<\tau_-$)  detector response reduces to a constrain  on the inductance L of the bias circuit. 
For given  bias conditions and intrinsic TES properties, the system is underdamped  for $L_{crit-}<L<L_{crit+}$ and the detector shows an oscillating response. The critical inductance $L_{crit\pm}$ is defined in \cite{IrwinHilton}.
Typically, the TES calorimeter read-out circuit is dimensioned such that the electrical circuit is faster than the thermal circuit with $L\le L_{crit-}$ where the best temperature-to-current responsivity is achieved. 
 In the limit of strong voltage bias $r_{sh}\ll R_0$ and high feedback gain   $\mathcal{L}_0\gg 1,\beta$, the critical inductance $L_{crit\pm}$ reduces to
 \beq{Lcrit_red}
 L_{crit\pm}\simeq \frac{CT_0R_0}{P_0}\left(\frac{3+\beta\pm 2\sqrt{2+\beta}}{\alpha}\right),
 \eeq
 which is defined solely  by the TES calorimeter physical parameters.

The form of the stability space is more complicated when a two-body thermal body is present \cite{Bennett_2010} or when the magnetic field dependence of the resistive transition needs to be included \cite{Sadleir_2014} and a more careful analysis is required.
 
The effect of the inductance L of the electrical bias circuit on the detector linearity in the small and large signal limit is further discussed in Section \ref{Large_signal} and shown in \figref{fig:pulses_L}.

\subsection{Fundamental Noise Sources}
\label{sec:noise}
\begin{figure}
\center
\includegraphics[width=9cm]{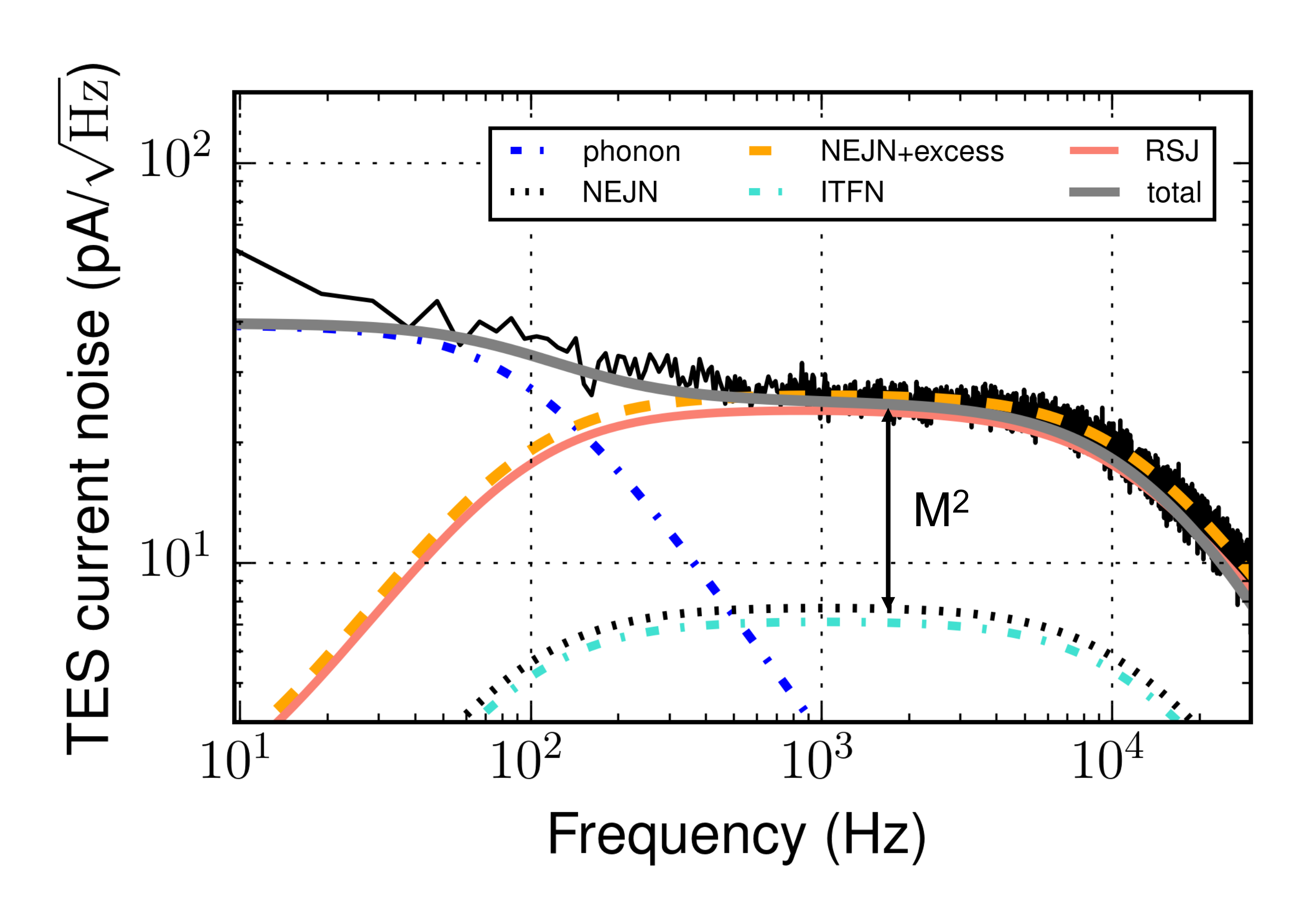}
\caption{\label{fig:schemeivnoise} Example of a current noise spectrum for a TiAu $80\times 20 \, \mu \mathrm{m}^2$ TES AC biased at 2.6 MHz, $R/R_N=13\%$ and $T_{bath}=55\,\mathrm{mK}$. The  different lines show the  noise contributions discussed in the text. The vertical arrow indicates the measured excess noise with respect to the estimated non-equilibrium Johnson noise (NEJN). The red solid line is the prediction of the Johnson noise based on the resistively shunted junction model. Reproduced from \cite{Gottardi2021}.}
\end{figure}

Noise sources in a TES-based microcalorimeter can be placed into three categories:
(1) noise sources internal to the TES, (2) noise sources from the circuit
in which the TES is embedded, such as Johnson noise in the bias resistor
and the noise contribution of the readout amplifier, and (3) noise
from the external environment such as magnetic field, RF-pickup, stray photon arrivals,
and fluctuations in the temperature bath.  The read-out amplification chain is typically dimensioned to give a negligible noise contribution on the detector performance, while the noise from the external environment  is generally  minimized by a proper design of the focal plane assembly.
The major intrinsic noise sources in a TES calorimeter originate from the thermal fluctuations in the phonon and electronic ensemble  of the device. Three intrinsic noise contributions are generally identified: the phonon noise, the internal thermal fluctuation noise and  the  Johnson-Nyquist noise.
\noindent As an example, we show in \figref{fig:schemeivnoise}, the typical current noise spectrum of a TES calorimeter under  AC bias \cite{Gottardi2021}.

The {\it phonon noise}, blue dot-dashed line in \figref{fig:schemeivnoise}, arises from the statistical energy  fluctuations generated during the random energy  exchange  between the TES-absorber body and the heat bath. The power spectral density of this noise is  $S_{ph}=4k_BT^2G_{bath}F(T_0,T_{bath})$, where $G_{bath}$ is the thermal conductance to the bath at a temperature $T_{bath}$. For thermal transport where the mean free path of the energy carriers is larger than the length of the link (radiative transport), $F(T_0,T_{bath})=((T_{bath}/T)^{n+2}+1)/2\sim 0.5$, \cite{Boyle59}, with  $n$  the thermal-process dependent  exponent. 
This noise is dominant at low frequencies in the detector thermal bandwidth typically below $\sim 200\mathrm{Hz}$.  

The {\it internal thermal fluctuation noise} (ITFN), cyan dot-dashed line, is generated by thermal fluctuation between distributed heat capacities internal to the TES-absorber, as for example in the two-body model shown in \figref{fig:ETscheme}. It has a power spectral density $S_{itfn}=4k_BT^2G_{12}$, where $G_{12}$ is the  thermal conductance between the two thermal bodies.  The  ITFN contribution can be derived from a proper characterization of the thermal circuit \cite{Hoevers00,Takei08,Kinn12,Maasilta12,Wake2019,Wake2020,deWit2021}.

The third main contribution to the TES total noise is the {\it Johnson-Nyquist noise} (JN) of the TES biased in the resistive transition. It can be written  in the form of a voltage noise $e_{int}=\sqrt{4k_BTR\zeta(I)}$. The~response of the TES current to $e_{int}$ is suppressed  at low frequency by the electrothermal feedback~\cite{Irwin1995} and becomes significant in the detector electrical band at kHz.The function $\zeta(I)$, which is strongly dependent on the TES current, conveniently describes the noise originated from the non-linear TES resistance nearly at equilibrium at the quiescent temperature $T_{0}$. It takes into account the non-linear correction terms to the linear equilibrium Johnson noise. For~a linear resistance, $\zeta(I)=1$. For a TES, the form of $\zeta(I)$  is not straightforward to model due to the non-linear and non-equilibrium  physical processes involved. 
The~characterization of the noise in the electrical bandwidth is complicated by the fact that both the Nyquist-Johnson noise and the internal thermal fluctuation noise  give a similar contribution  in the measured TES current noise,  after~passing through the system~transadmittance \cite{Maasilta12,Wake2018,Wake2019}. 
To calculate $\zeta(I)$, Irwin~\cite{Irwin2006} assumed the TES to be a simple Markovian system with no hidden variables such as internal temperature gradients and fluctuating current paths. By applying the Stratonovich’s nonequilibrium Markovian   
fluctuation–dissipation relations, he calculated the first order, near-equilibrium,  non-linear correction term to the noise to be $\zeta_{ne}(I)=1+2\beta$ \cite{Irwin2006}. 

In~this form, the~noise model is known as the non-equilibrium Johnson noise  and it  is extensively used in the literature  to model the, $\beta$ dependent, Nyquist--Johnson voltage noise observed in the TES. However, the~broadband noise,  typically observed in the TES electrical bandwidth \cite{Ullom04}, could only partially be explained after the introduction of the correction term $\zeta_{ne}(I)$ and only  at relatively low $\beta$ values~\cite{Takei08,Jethava09,Smith13}.
Smith~et~al.~\cite{Smith13}, for example, observed unexplained noise at high frequency after excluding the presence of a significant contribution of the ITFN  in their data taken with  low resistance, large $\beta$, high thermal conductance devices. 

New experimental results on the noise characterization of TES microcalorimeter \cite{Gottardi2021,Wessels2021} indicate that the observed Nyquist--Johnson voltage could find a theoretical explanation in the framework of the Josephson effects in the superconducting film described in Section~\ref{sec:TESPhysics}. Due to the non-linear  current–to-voltage relationship of the TES, the thermal fluctuation noise of the oscillating current induced at the Josephson frequency, much higher than the detector bandwidth,  is mixed down and leads to an increase of noise in the TES electrical bandwidth (red solid line labelled RSJ in \figref{fig:schemeivnoise}). This noise mechanism will be discussed further in Section~\ref{sec:RSJnoise}.

A detailed analysis of the thermometer sensitivity, the~detector noise and the signal bandwidth, shows that the minimum energy resolution achievable with a TES calorimeter can be written  as  
\begin{equation}\label{eq:dEnoise}
\Delta E_{FWHM}\simeq 2\sqrt{2\ln 2}\sqrt{4k_BT^2_0C\frac{\sqrt{\zeta(I)}}{\alpha}\sqrt{\frac{nF(T_0,T_{bath})}{1-(T_{bath}/T_0)^n}}}.
\end{equation}
 The unit-less parameter  $F(T_0,T_{bath})$ and the thermal conductance exponent $n$  depend on the physical nature of the thermal conductance and whether,  for example, the phonon transport from the TES to the heat sink  is radiative or diffusive \cite{McCammon2005}. For~a TES, $F(T_0,T_{bath})\simeq 0.5$ and $n\simeq 3$ \cite{IrwinHilton}.  
 To~achieve  the ultimate sensitivity, for~given $C$ and $T_0$, the~factor $\sqrt{\zeta(I)}/\alpha$ has to be minimized. 

 \subsection{Non-linearity} \label{Large_signal}

The equations and discussion outlined in Section \ref{Basic_principles} to \ref{sec:noise} describe the TES behaviour in the small signal limit. This assumes the TES response as a function of photon energy is purely linear. TESs however, are generally non-linear devices and the pulse shapes (and noise properties during the pulse) will vary depending upon the deposited energy. Non-linearity comes about from two basic effects. Firstly, the shape of the of resistive transition $R(T,I)$ is inherently non-linear (see for example \figref{fig:RTI}). The temperature increase of the TES is proportional to the photon energy ($\Delta T = E/C$). The larger the $\Delta T$, the closer the TES resistance is driven towards the normal branch of the transition and eventually the response of the TES will saturate as $R_0+\Delta R$ $\rightarrow$ $R_n$. Secondly, even for a linear resistive transition (constant dR/dT and dR/dI), non-linearity occurs due to the voltage biased nature of the electrical bias circuit itself. The maximum current change from the X-ray event $\Delta I$ is constrained by the TES equilibrium bias current $I_0$, thus the current response will gradually saturate as $\Delta I$ $\rightarrow$ $I_0$. The TES current scales roughly as $1/R$, thus for every doubling of $\Delta R$, the corresponding $\Delta I$ will be reduced by a factor of 2. The current and resistance non-linearity will both contribute to the large signal behaviour of TES detectors. Non-linearity not only affects the pulse shapes, but it also changes the intrinsic detector noise during the X-ray pulse. Generally, as the resistance increases the total detector noise will decrease, this can have a beneficial effect of compensating for the decreased amplitude of the pulse shape due to the non-linearity. The combined effect on the energy resolution is dependent upon the pulse processing method, the details of which will be discussed in Section \ref{Pulse_processing}.

Multiplexed readout schemes such as time-division-multiplexing (TDM) and frequency-domain-multiplexing (FDM) only have a finite available bandwidth (see Chapter "Signal readout for Transition-Edge Sensor X-ray imaging spectrometers" for a detailed discussion on these readout schemes). In order to optimally match the bandwidth of the detectors to the readout, a circuit inductance $L$ is typically chosen to critically damp the detector response. In the small signal limit, adding circuit inductance changes the detector responsivity, but does not affect the fundamental energy resolution (the signal and noise are affected equally). However it is interesting to consider that the increase in inductance can affect the detector linearity in the larger signal limit.

\begin{figure}[h!]
\centering
  \includegraphics[width=1.0\textwidth]{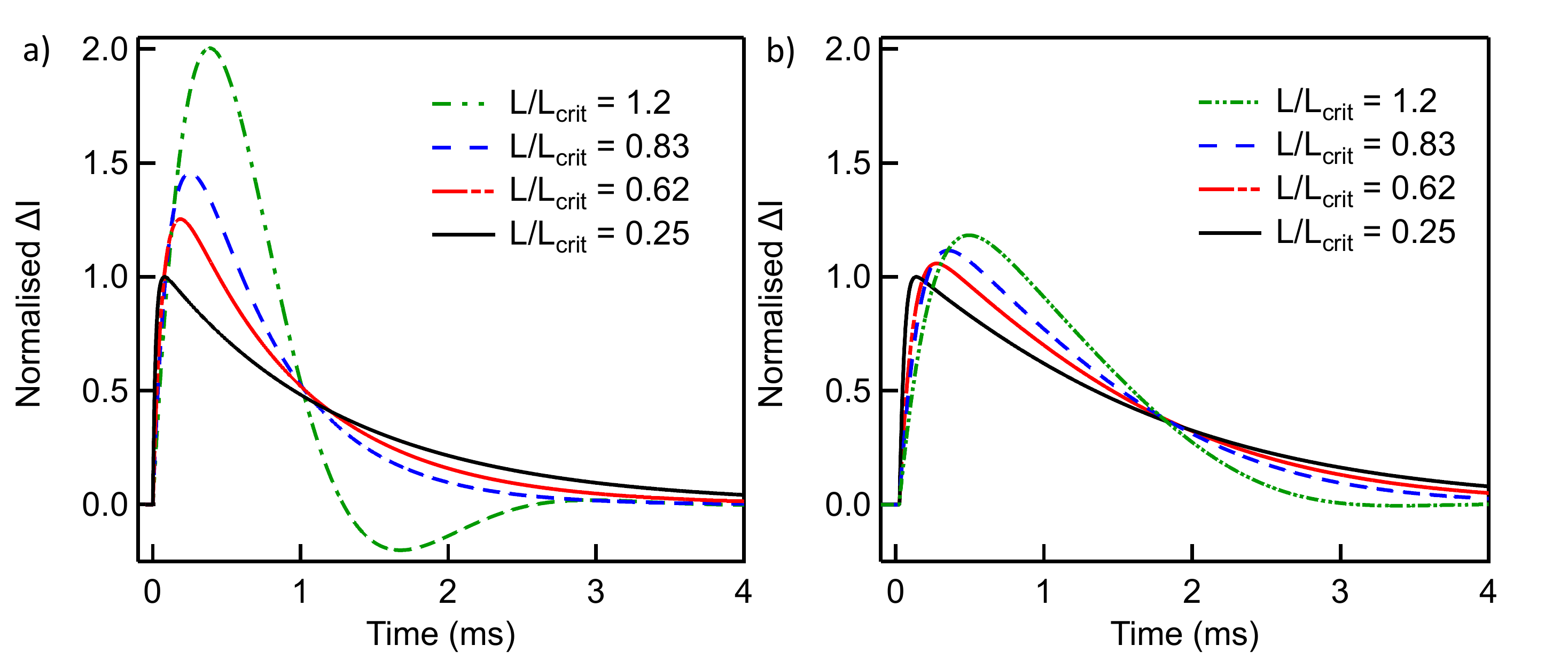}
  \caption{(a) Small signal modelled pulse shapes as a function of $L/L_{crit}$. (b) Measured pulse shapes at 6 keV as a function of $L/L_{crit}$. Data are normalized to the peak of the low inductance pulse-height in both cases. These data show how the small signal pulse height is strongly affected by $L/L_{crit}$, whereas due to the effects on non-linearity the measured large signal data shows only a modest increase in pulse-height.}
  \label{fig:pulses_L}
\end{figure}

\figref{fig:pulses_L} a) shows the small signal modelled pulse shapes as a function of $L/L_{crit}$ for an Athena/X-IFU pixel compared to b) the measured pulse shapes at 6 keV as a function of $L/L_{crit}$ \cite{Smith2021}. Data are normalized to the peak of the low inductance pulse-height in both cases. These data show how the small signal (linear) modelled pulse shape is strongly affected by $L/L_{crit}$. The TES response is not limited by either current or resistance non-linearity, and the pulses become larger and faster as $L/L_{crit}$ is increased. Whereas for the real device, which has a non-linear response to 6 keV X-rays, the measured large signal data shows only a modest increase in pulse-height with increasing $L/L_{crit}$. This is because the inherent non-linear response of the real device suppresses the increased current and resistance changes predicted from the small signal model. Thus the current response of the real device becomes increasing more nonlinear as a function of energy, with increasing $L/L_{crit}$. Significant non-linearity may impact the achievable resolution of the device and complicate the energy scale calibration. Thus for some device designs and applications it may be preferable to operate below the small signal critically damped limit of $L/L_{crit}$ = 1.

\subsection{Pulse Processing} \label{Pulse_processing}

The output signal of a TES calorimeter is typically the amplified and digitised change in the TES current.  The time stream of the detector output is sampled into records of a fixed length, which can be represented as a vector $\bf S$. In the small signal limit, and for records containing  only a single X-ray pulse of energy $E_{k}$, the signal is proportional to the energy ${\bf S}=E_k{\bf T_{k}}+{\bf n}$, where $\bf T_{k}$ is the noise-free signal and $n$ is the stationary additive noise from the detector and electronics.    
The best estimate of $E_k$, when the noise is not-white, would be given by minimising, with respect to $E_k{\bf T}_k$, the $\chi^2$:
\beq{chi2}
\chi_k^2=({\bf S}-E_k{\bf T}_k)^T{\bf W}({\bf S}-E_k{\bf T}_k),
\eeq
being ${\bf W}={\bf N}^{-1}$, the inverse of the noise variance-covariance matrix, where the diagonal elements are the variances, $N_{ii}=\sigma_i^2$ and the off-diagonal elements are the covariances, $A_{ij}=\sigma_{ij}$ from the noise sources. 
This leads to the following definition for the pulse energy
\beq{Ek}
E_k= {\bf S}^T\left[\frac{{\bf W}{\bf T}_k}{{\bf T}_k^T{\bf W}{\bf T}_k}\right]={\bf S}^T{\bf F},
\eeq
where $\bf F$ is the optimal filter. First introduced by Szymkowiak et al. (1993)  \cite{Szymk1993}, the optimal filtering technique has been used routinely in the analysis of X-ray micro-calorimeter pulses  in laboratory experiments \cite{Whitford2005,Fowler2016,Fowler2017} and on  space satellites \cite{Boyce1999,Seta2012,Tsujimoto2018}.

In practice, the optimal filter $\bf F$ is constructed from a model of the pulse shape and the noise. The noiseless signal ${\bf T}_k$ is obtained  by averaging together many records with pulses of a known, identical energy.   Depending on whether the filter is build in the time \cite{Fixsen2004}  or frequency \cite{Szymk1993} space, the noise is defined by the noise autocorrelation function or  by the power spectral density estimated from the pulse-free records.
If the noise is stationary, ${\bf WT}_k$ is  a convolution or  a simple product  in the time or frequency space, respectively.  

This pulse analysis is optimal under the assumption that \cite{Whitford2005, Fowler2016}: (1) the detector response is linear, (2) the pulse shape for a specific energy is known, (3) the noise is stationary and the autocorrelation function is known, (4) the noise is additive and follows a multivariate  Gaussian distribution, (5) the pulses have been acquired while the sensor is in its steady-state condition and the energy from earlier X-rays has been fully dissipated.

In the energy range where the detector response is linear, the pulse shape estimated at one energy could be used to construct the optimal filter for all the energies. In practice, this is never the case and an energy scale calibration program is always required to build an optimal filters look-up table to accurately estimate the incoming photon energy.

Several approaches has been proposed  to deal with the  detector
non-linearity, the  non-stationary noise  and the position dependency.  Those include interpolating the filters for an arbitrary energy \cite{Whitford2005,Shank2014} , Taylor expansion of a continuous
pulse shape model to leading order in energy \cite{Fowler2017}, position dependent analysis \cite{Smith2006}, calculating the energy
dependence of the response of the detector operated in electrothermal feedback mode \cite{Hollerith2006,Fowler_2018}, multi-pulse fitting \cite{Fowler_2015}, and principal component analysis \cite{deVries2016,Fowler2020,Busch2016}. 

When constructing the optimal filter, the time average of the optimally filtered pulse, or  equivalently, the zero frequency bin of the filter calculated in the frequency space, is generally ignored. The zero frequency bin however, contains crucial information on slow varying fluctuations at the detector and the read-out system level and discarding it might result in a signal-to-noise reduction. 
As it was shown by  Doriese et al. \cite{Doriese2009}, since the size of the zero-frequency bin is determined by the pulse record-length, the energy resolution
of a TES depends as well on the record length. An optimal record length does exist, which is typically defined by the detector internal time constants, the count-rate and the computational resources. It is however possible to at least partially recover the energy resolution degradation by correcting the baseline signal after linear fitting the data before and after the pulse. A post-filtering correlation analysis of the filtered pulse-height and the detector baseline level is routinely used as well to mitigate the effect of slow detector drifts occurring during the pulse acquisitions.
 
The photon flux experienced by a single pixel during bright sources observation could be so high that it becomes impossible to  isolate each single photon and perform optimal filtering. Defocusing techniques could be used to spread the high X-ray flux  over a large number of pixels in the array. On the ESA's Athena mission for example, a defocusing of 35 mm (with respect to the nominal focal length of 12 m) will enable the observations of extremely bright galactic sources with fluxes up to $\sim 1$ Crab, with only limited
spectral resolution degradation \cite{Peille2018,Kammoun2022}.

\begin{figure}
\center
\includegraphics[width=0.75\textwidth]{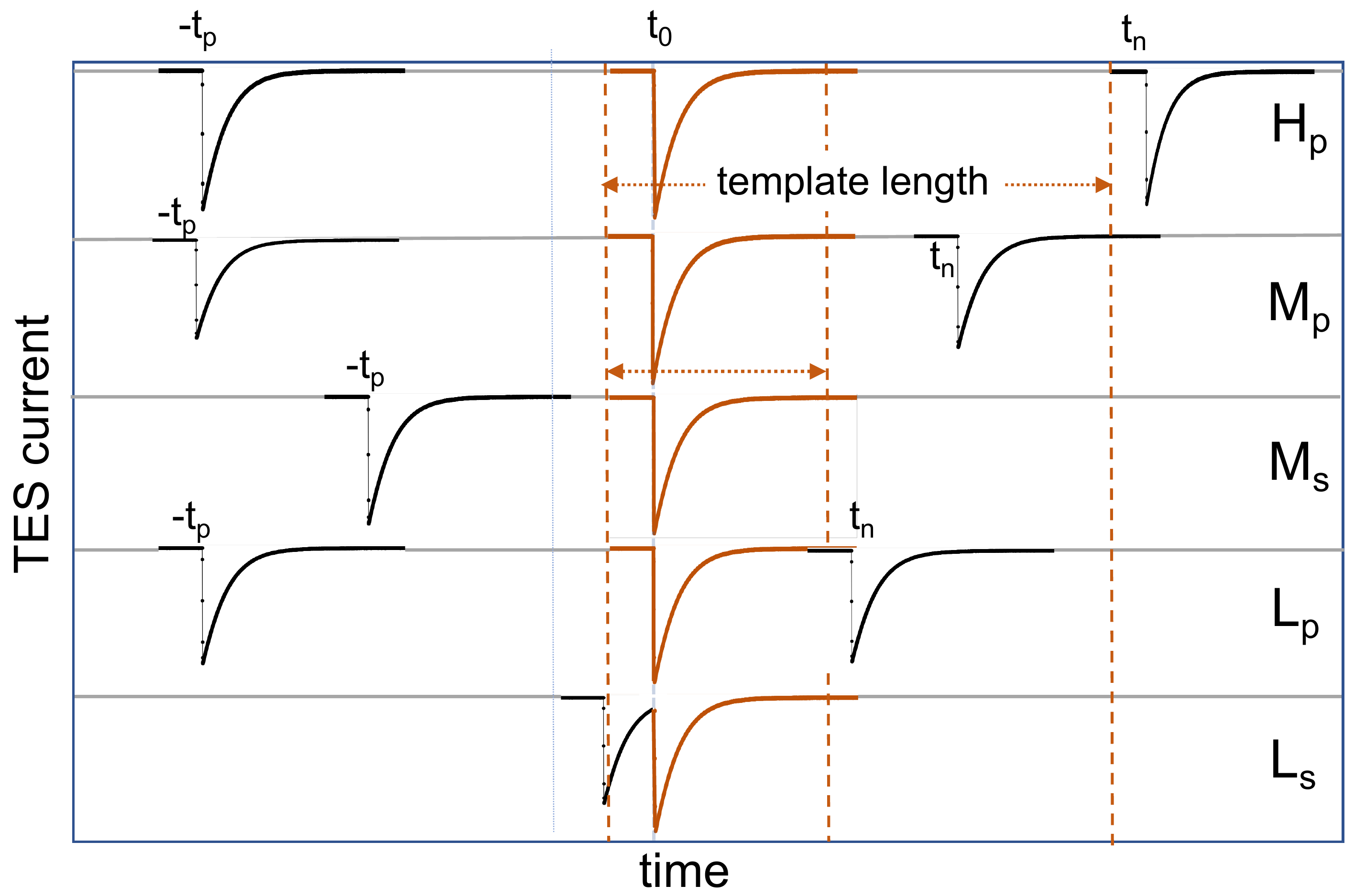}
\caption{\label{grading}Definition of grade for pulse, where $t_p$ is the time since previous pulse and $t_n$ is the time until next pulse. The labels  $H_p$,  $M_p$,  or  $L_p$ indicate the high-,medium- and low-resolution primary events,  respectively.  The  events  following  the primary  $M_p$  or  $L_p$  events  are  labelled  as secondary with an $s$ suffix \cite{Takeda2014}.}
\end{figure}

On a space telescope, most of the  data are processed on board. The main on-board processes consist of the trigger algorithm, the event reconstruction algorithm to estimate in real time the arrival time and pulse energy, and  the filter generation to compute the template and the optimal filter. The energy scale and the energy scale calibration  processes performed on ground  shall convert the biased energy estimate into calibrated units, by using fiducial photons from calibration sources. Given the limited amount of computational resources and the limited transmission rate of the satellite telemetry, only a small amount of information is transmitted to Earth. At high photon fluxes, a good strategy is required to estimate the pulse heights with high resolution in the presence of event pile-ups. X-ray instruments on board of the Astro-E \cite{Boyce1999} and  Hitomi (Astro-H) \cite{AstroH2018} introduced the {\it event grading} approach as illustrated in \figref{grading}.
The X-IFU instrument shall follow a similar grading definition. The event grading algorithm qualifies the detected pulses according to the proximity of other events in the same record. On X-IFU, High resolution events corresponds to isolated pulses not affected either by preceding or succeeding ones and have a resolution of 2.5 eV at 7 keV. Medium and Low resolution events have a resolution of $> 2.5$ and $>3$ eV, respectively, degraded by another pulse arriving too early after the main one. \figref{grading} shows
the definitions of the grades used on Hitomi.  A $p$ suffix denotes that the current event 
occurred at least $n$ samples  after the previous event. When 
the interval before the next event is more than $n$ , between $m$ and $n$, or less than $m$ samples, the current 
event  is  designated  $H_p$,  $M_p$,  or  $L_p$ (high/medium/low-resolution),  respectively.  The  events  following  the primary  $M_p$  or  $L_p$  event  are  labelled  as secondary with an $s$ suffix \cite{Takeda2014}.

\section{Detector Design}\label{Detector_design}
\subsection{TES Properties }\label{TES_design}

The TES is the core of an X-ray microcalorimeter. Various types of superconducting films have been successfully used for different applications: a single layer of tungsten~\cite{W_CRESST2018,W_Lita05}, alloys such as Al-Mn~\cite{AlMn2011}, and proximity-coupled bilayers (or multilayers) made of Ti/Au~\cite{Nagayoshi19}, Mo/Au~\cite{Fink2017,Parra2013,Mo_Fabrega2009,Cherv2008}, Mo/Cu~\cite{Orlando18}, Ir/Au~\cite{IrAu_Kunieda2006}, and Al/Ti~\cite{AlTi_Lolli2016,AlTi_Kamal2019}. These different TES films can all be designed to achieve transition temperatures of 50-100 mK, which is essential for high resolution spectroscopy. The superconducting leads connecting the TES to the rest of the bias circuit must have a higher transition temperature and critical current with respect to the TES film. Niobium is widely used for the leads, but~other superconductors such as niobium nitride, molybdenum, and aluminum are also applied.

The TES pixel geometry as well as the design of the coupling between the TES bilayer and the X-ray absorber has direct impact on various properties of the detector. This includes the thermal conductance to the bath \cite{Hoevers2005,Hays2016,Morgan2017,deWit_2020}, the TES transition smoothness and uniformity \cite{Smith13,Zhang2017}, the susceptibility to the external magnetic field,  the detector normal state resistance \cite{Gottardi17,Sakai2018,Morgan2019,deWit_2020}, and the overall detector noise \cite{Ullom04,Sadleir11,Miniussi2018,Wake2018,deWit2021}. In particular, the pixel design optimization is different depending on whether the X-ray microcalorimeters are DC or AC biased \cite{GottNaga2021}. Additional normal metal layers such as 'banks' that run down the edges of the device, parallel to the direction of current flow, and 'zebra-stripes', that run perpendicular to the current flow, were widely explored in early TES design for empirically controlling the transition shape and noise properties \cite{Ullom04,Lind_NIMA_2004}. Whereas the use of zebra-stripes has been found to be beneficial to some designs, they can introduce additional inhomogeneity into the film properties (non-uniform current and magnetic field distribution), which greatly complicates the device physics. This can be detrimental to device performance, particularly reproducibility and uniformity in large format arrays \cite{Smith13,Smith2014,Wake2018,Zhang2017}. Thus, these types of features are not universally adopted on all TES devices \cite{Miniussi2018,Smith2020}. Understanding all these geometry dependent effects on TES performance has been an a very active area of research and has enabled better control of the device properties and ever improving energy resolution. 

\subsection{Thermal Isolation}
\label{Thermal_isolation}
If the thermal conductance ($G_{bath}$) from the sensor to the heat-bath is too great, the pixel response time may be so fast that it becomes impractical to measure the electrical pulses with the available readout bandwidth. For this reason, it is important to limit the thermal coupling to tune the speed of the detector to meet both the count-rate requirements of the application and bandwidth requirements of the readout. The conductance to the heat-bath depends on the material and geometry of the sensor \cite{Morgan2017,deWit_2020} and the substrate on which it is deposited. Thermal control is often achieved by isolating the TES on thin SiN (or SOI) membranes which are typically 100 nm to a few $\mu$m's thick. In the 2-D ballistic phonon transport limit $G_{bath}$ will scale with the perimeter of the TES in contact with the membrane and scale as $\sim T^{2}$ \cite{Stahle2002,Hoevers2005}. Additional thermal control has been demonstrated by engineering slots in to the membrane to reduce the phonon emission and thus the conductance to the bath. Metalization layers can also be added to the surface of the membrane to either enhance \cite{Hays2016} or suppress \cite{Zhang2019} the phonon transport without changing the geometry of the TES itself. TESs can also be directly fabricated on bulk Si substrates. Without the isolating membrane $G_{bath}$ will be limited by the acoustic mismatch (Kapitza boundary resistance) between the TES and the substrate, and/or electron phonon decoupling \cite{Wellstood1994}. The $G_{bath}$ from acoustic mismatch will scale as the area of the TES in contact with the substrate and as $T^{3}$. Whereas electron-phonon decoupling scales as the volume of the sensor and $T^{5}$ (or $T^{6}$). Whichever term dominates will depend upon the geometric details of the devices and the temperature of operation.

\subsection{Absorber Design and Properties} \label{Absorbers}
For X-ray  applications where high density, high fill-factor arrays are required, the absorbers are typically cantilevered above the TES and underlying membrane, making thermal contact with the TES via support columns. The absorber needs to rapidly thermalize the photon energy and conduct it to the TES. If the thermal diffusion from the absorber to the sensor is insufficient, the pixel may exhibit excess thermal noise and position dependent resolution broadening (which occurs when the measured signal at the TES depends on the photon absorption position). The absorber material may also contribute significantly to the total heat-capacity, which is an important design parameter used to tune the desired pixel energy range and resolution. Thus, the composition and geometry of the absorber must be optimized to provide the necessary thermal diffusion and heat-capacity, whilst also achieving the desired vertical quantum efficiency (stopping power) for incident X-rays. High-Z metals such as Au are ideal because they have short X-ray attenuation lengths and thermalize quickly. However, the large carrier density means they also have a relatively high specific heat, making them less ideal for applications requiring large area, high stopping power and high energy resolution. The semi-metal Bi is also an interesting candidate material because it has an intrinsically low carrier density and specific heat. However, the thermalization properties of Bi are highly dependent on the deposition method. TESs with evaporated Bi absorbers have shown non-Gaussian broadening in the measured X-ray spectral response (commonly referred to as ‘low energy tails’). This is not fully understood, but may be related to the trapping of energy in long lived states. Electroplated films tend to have larger grains than than in evaporated films and can achieve a higher Residual-Resistivity Ratio (RRR), defined as the ratio of the resistivity of a material at room temperature and at 4 K, and better thermal characteristics \cite{Yan17,Brown08}. Electroplated absorbers have shown the ability to achieve high resolution spectra without significant spectral broadening features \cite{bandler2008,Brown08}. For applications where large area absorbers are required, composite Bi/Au absorbers can be used to combine the best properties of both materials. This is the approach for the Athena/X-IFU detectors where a bottom 1 $\mu$m Au layer provides the desired heat-capacity and rapid thermal diffusion to the sensor. A 5 $\mu$m Bi layer is added on top, which provides additional stopping power without adding significant additional heat-capacity. These designs provide $>$ 90 $\%$ vertical quantum efficiency at 7 keV combined with $>$ 96 $\%$ filling-factor. To minimize the effect of the absorption of the stray light in the detector, it is also desirable to increase the reflectivity of the absorber for long-wavelength infrared radiation.  For this purpose, a  Au capping layer on top of the Bi can be added. For pixels being developed for the Athena/X-IFU instrument, an increase in reflectivity, from~45$\%$ to 80$\%$, has been  measured at room temperature after adding a 40 nm Au capping~layer~\cite{Hummatov2020}.

\subsection{Current State of the Art} \label{state_of_art}
\begin{figure}
\center
\includegraphics[width=1\textwidth]{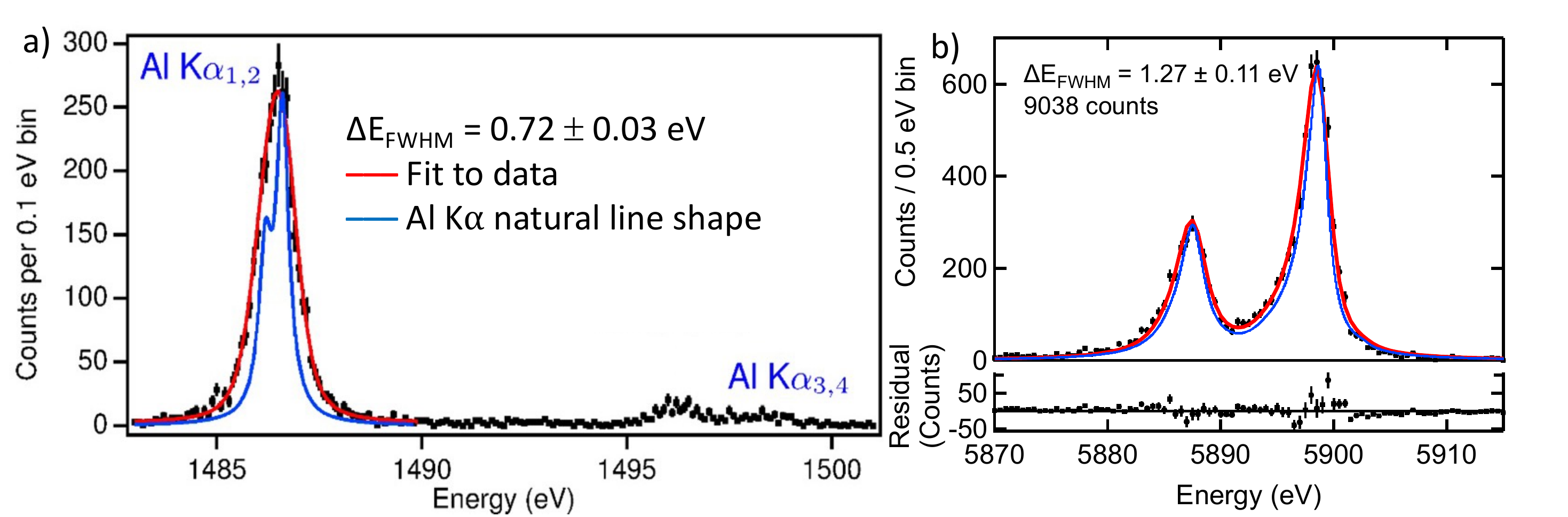}
\caption{\label{fig:spectra} Best achieved spectral resolution in TES microcalorimeters at an energy of a) 1.5 keV (Al-K$\alpha$) \cite{Lee2015} and b) 5.9 keV (Mn-K$\alpha$) \cite{Sakai2019}. The blue lines are the natural line shapes of the line complexes and red lines are the fit to data. These are both DC-biased Mo/Au bilayer TESs developed at NASA/GSFC.}
\end{figure}

TES detectors for X-ray astrophysics applications have now achieved exquisite resolution performance over a broad range of energies. At an energy of 5.9 keV (Mn-K$\alpha$) TES microcalorimeters can routinely achieve below 2 eV. DC-biased devices developed for various mission concepts at NASA/GSFC, using Mo/Au bilayers and either Au or Au/Bi absorbers, have demonstrated around 1.6 eV FWHM at 5.9 keV (Mn-K$\alpha$) \cite{Miniussi2018,Busch2016,Smith2012}. With the best energy resolution performance to date being 1.3 eV FWHM \cite{Sakai2019}. This was achieved on a 12 $\mu$m TES with $31\times 31 \, \mu\mathrm{m}^2$, 4 $\mu$m thick Au absorber. This very small pixel design is being developed of high angular resolution applications such as NASA's Lynx \cite{Bandler2019}. At an energy of 1.5 keV (Al-K$\alpha$), sub-eV resolution has now been achieved several times with a best resolution of 0.7 eV FWHM \cite{Lee2015} (see \figref{fig:spectra}).

\begin{figure}
\center
\includegraphics[width=1\textwidth]{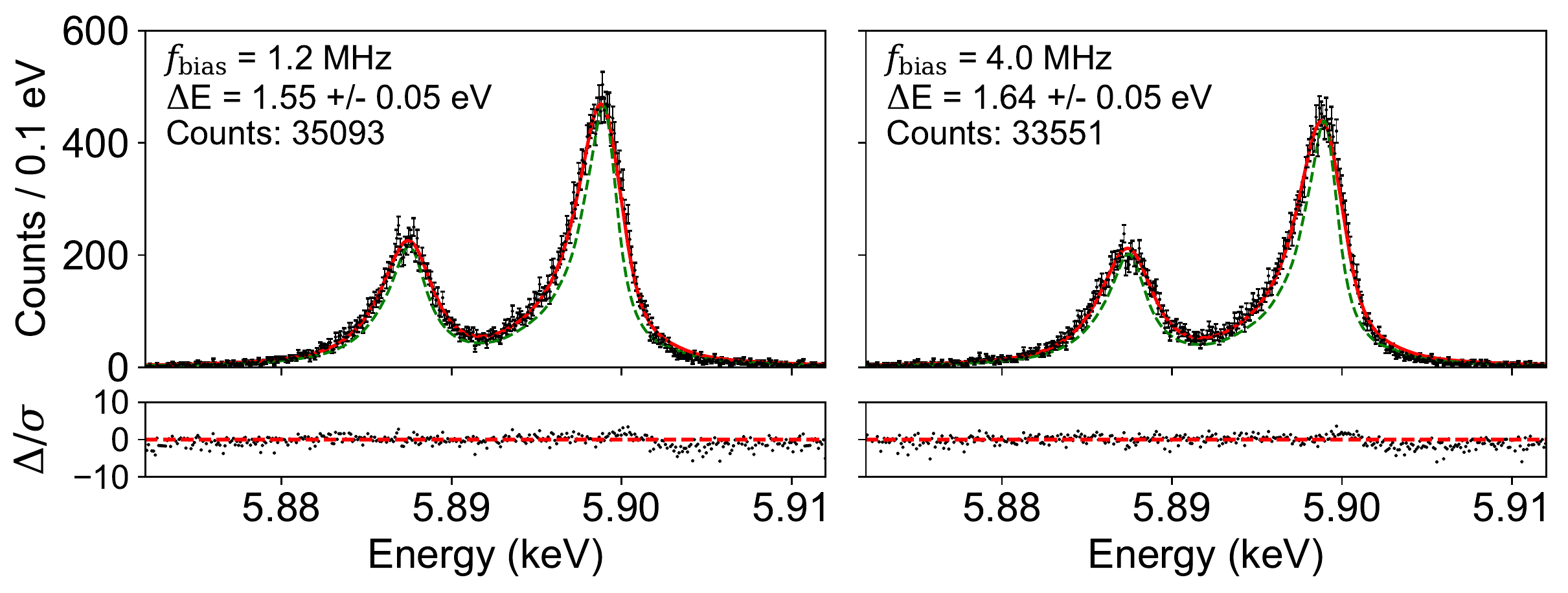}
\caption{\label{fig:ac_spectra} Best achieved spectral resolution at 5.9 keV (Mn-K$\alpha$) with Athena/X-IFU-like TiAu TES microcalorimeters developed at SRON and AC biased respectively at (left figure) 1.2 and (right figure) 4.0 MHz. The green-dashed line is the natural line shape of the Mn-K$\alpha$ complex. }
\end{figure}

Similar performance has been achieved recently under AC bias using pixels made of Ti/Au  TESs with a $250\times 250 \, \mu\mathrm{m}^2$, $2.35 \, \mu\mathrm{m}$ thick Au absorber, developed at SRON and optimized for the FDM readout of Athena/X-IFU-like instruments on future X-ray missions. In \figref{fig:ac_spectra}, the Mn-K$\alpha$ spectra  for two pixels, biased respectively at 1.2 and 4.0 MHz, are shown. The detectors have a $T_c\simeq 80 \,\mathrm{mK}$, heat capacity $C=0.76 \, \mathrm{pJ/K}$. normal resistance $R_N\simeq 200 \,\mathrm{m}\Omega$ and a thermal conductance to the bath $G_{bath}\simeq 50 \, \mathrm{pW/K}$.

These results are representative of the best single pixel spectra achieved to date in X-ray microcalorimeters. The expected performance of a large format array in a space flight instrument will, however, typically be somewhat worse than the best individual pixel results measured in a laboratory environment. This is due a variety of additional sources of noise associated with a large instrument and its environmental conditions. This includes thermal cross-talk between pixels in the array, electrical cross-talk in the bias and readout circuity, noise from the multiplexed readout electronics, drifts from the instrument environment, noise from particle background events, and noise from micro-vibrations. This means that the individual pixel resolution must be optimized to be lower than the final instrument requirement. For the Athena/X-IFU satellite instrument \cite{XIFU2018}, for example, the individual pixel resolution must be less than 2.1 eV at 7 keV in order to satisfy the full instrument requirement of 2.5 eV.

\section{Physics of the Superconducting Transition}
\label{sec:TESPhysics}

\subsection{The Superconducting Transition}

The superconducting transition of a TES,  since its invention,  has been described,  according to the well established microscopic (BCS theory \cite{BCS57}) and macroscopic theory of superconductivity in low critical temperature $T_{ci}$ material.  In particular, the macroscopic Ginzburg-Landau  theory \cite{GL1950} turned out to be rather successful in explaining the physics near the superconducting transition. 

However, while the physics of conventional superconductors is well understood, in two dimensional superconducting films driven in the transition by a large current, like TESs, the observed phenomena are more difficult to explain.   
For many years, the interpretation of the observed properties of the TES superconducting transition, such as $T_c$,  width $dT_c$, and steepness $dR/dT$, was given in the framework of large transport current, Joule heating, self-magnetic field, film impurities, phase slips effects, and vortices   generations and annihilations \cite{IrwinHilton}.
Interesting enough, researchers overlooked for many years the potential impact  on the transition shape of the higher $T_c$ superconducting leads connecting to the TES bilayer.
Superconducting structures, in the form of sandwiches or bridges, consisting of a combination of superconducting and normal metal films or superconducting films with different $T_{c}$  has been studied since the 70's in the framework of superconducting  junctions and weak links \cite{Likharev79}. 
It was first reported by Sadleir~et~al.~\cite{Sadleir10,Sadleir11} that TES structures behave as a superconducting weak-link due to the long-range longitudinal proximity effect originating from the superconducting Nb leads. Their conclusion was based on three solid experimental findings: (i) the exponential dependence of the critical current $I_c$ upon the TES length $L$ and the square root of the temperature difference $T-T_{ci}$, with $T_{ci}$ the intrinsic critical temperature of the bilayer, (ii) the scaling of the effective transition temperature $T_{c}$ and the transition width $\Delta T_c$ as $1/L^2$, and~(iii) the Fraunhofer-like oscillations of the critical current as a function of the applied perpendicular magnetic field.
We will discuss the details of these findings in the next section. 

The physics of superconducting junctions and weak-links  has been extensively treated  in many books and publications \cite{Likharev79,BarPat1982}. 
In 1962, Brian D. Josephson predicted the existence of several new phenomena in two superconductors weakly coupled to each other by a sufficiently thin insulating material \cite{Josephson1962}. Josephson argued that a supercurrent should leak through the thin barrier depending on the voltage across the junction following the law of quantum mechanics. Josephson's prediction was  experimentally verified few years later by P.W. Anderson and J.M Rowell \cite{AndRow1963}. These works should be considered as milestones for the starting of a new era for the superconducting electronics and quantum physics. Josephson's predictions turned out to be valid not only in superconducting structures, but in other macroscopic systems as well. The Josephson effects have been observed for example in the beautiful experiments with superfluid He \cite{DavisPack2002} and Bose-Einstein condensates in trapped-atom  interferometers \cite{Levy2007}.
\begin{figure}
\center
\includegraphics[width=11cm]{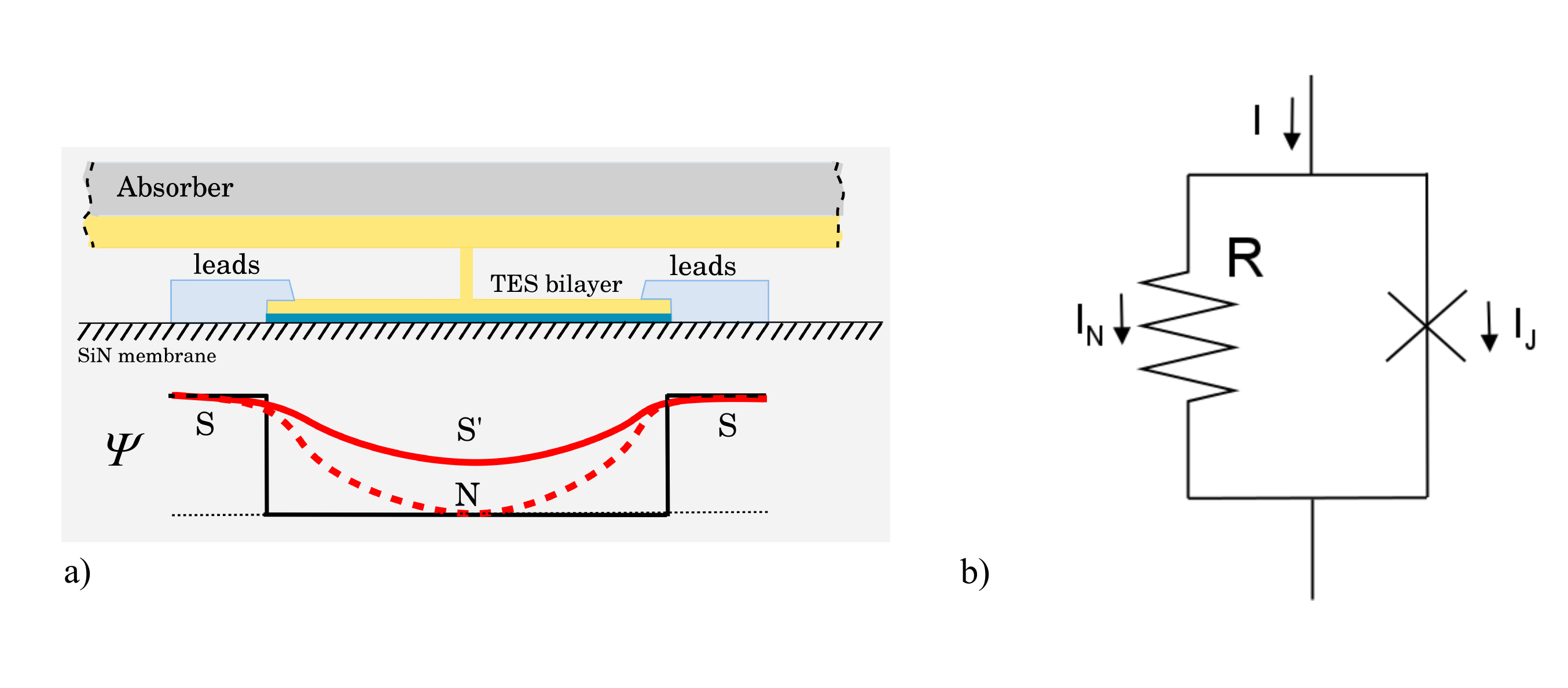}
\caption{\label{fig:WLRSJ} a) Schematic representation of a TES-lead system as a superconducting weak-link. The dashed area is the silicon nitrite membrane, the bluish layer is the TES superconducting film (Ti,Mo,..), the yellow layer is the Au film of the TES-absorber system and the grey area represents the Bi layer on top of the absorber. As shown below the drawing, in~the proximity of the leads, the spatially varying superconducting order parameters $\Psi$ of the TES bilayer is enhanced. b) The RSJ model.}
\end{figure}

The phenomena described by Josephson are summarized in two fundamental equations, which describe respectively a DC and an AC effect:
\beq{dcJosephson}
I_J(\varphi)=I_c\sin \varphi
\eeq
and 
\beq{acJosephson}
V(t)=\frac{\hbar}{2e}\frac{\partial \varphi}{\partial t}.
\eeq
According to the first equation (dc Josephson effect) the superconducting current has a  sinusoidal dependence to the phase difference of the weak-link, i.e. the current through the weak-link is zero when the phase difference is zero, or a multiple of $\pi$, and $I(\varphi)$ is a periodic function with period $2\pi$, such that a change of the current flow  direction  will cause a change of sign in $\varphi$.
The second equation (ac Josephson effect) states that  the rate at which the phase difference $\varphi$ changes in time is proportional to the voltage across the  weak-link.
Both \Eqsref{dcJosephson}{acJosephson} can be derived using a straightforward quantum-mechanics formalism describing a two-energy-level system, as it was done for example by Feynman in one of his lectures \cite{Feynman:1957:SS}. When the junction is a narrow bridge or a thin normal metal film, the dc Josephson equation (\Eqref{dcJosephson}) can be derived following a different approach as described by Aslamazov and Larkin \cite{AslLar1969}.  They showed that, for weak-links with length $L$  smaller than the superconductors coherence length $\xi$
 ($L\ll\xi$), the supercurrent density through the weak-link can be derived from the two Ginzburg-Landau equations  as a result of interference of  the wavefunctions of the two coupled superconductors. 

If the current applied to the junction (or weak-link) exceeds the critical value $I_c$, dissipation occurs and a normal electron or quasiparticle current $I_N$ starts flowing through the junction as well, in parallel with the superconducting Josephson current $I_J$.
The total current $I$ in a junction, with  normal-state resistance $R$, is then the sum of the normal current $V/R$ and the Josephson current $I_J=I_c \sin\varphi$:
\beq{eq:RSJ}
I=I_c\sin\varphi+\frac{\hbar}{2eR}\frac{\partial \varphi}{\partial t}.
\eeq
\Eqref{eq:RSJ} is the basic equation of the so-called resistively shunted junction model (RSJ), which describes the system as an electrical circuit consisting   of a  Josephson junction in parallel with a normal resistance (\figref{fig:WLRSJ}.b).

A typical  TES, fabricated from a bilayer with intrinsic critical temperature, $T_{ci}$,  smaller than the critical temperature of the leads $T_{cL}$, can be treated as a superconducting $SNS$ or $SS'S$ weak-link \cite{Likharev79,Golubov04}, with the bilayer being $N$ when $T>T_{ci}$ and $S'$ when $T<T_{ci}$.
 
In some devices however, the Josephson effects described above become less evident and tend to disappear.  The TES operates in a stronger superconducting regime, depending on the exact shape of the $I_c(T)$, the bias current and the operating temperature. As the temperature is further reduced to $T<<T_{ci}$, the measured critical current follows the behavior typical of the Meissner state for a strongly coupled superconductor. 
Before the discovery of weak-link effects in TES's, Irwin et al. \cite{Irwin2018} proposed  a simple two-fluid model to describe the total current through a TES biased in the resistive transition
\beq{twofluid}
I_{TES}(T ) = c_I I_c(T )+ \frac{V_{TES}}{c_RR_N}
\eeq
This model is a simplified form of the Skocpol-Beasley-Tinkham (SBT) model [7] of a resistive transition
for a superconductor with phase slip lines (PSL) assuming a constant ratio of the time-averaged critical current to the critical current.
The model defines a supercurrent as some fraction ($c_I$) of the critical current, and a quasiparticle current, which
is equal to $V_{TES}$ divided by some fraction ($c_R$) of the normal resistance. Although it does not provide a physical mechanism
for the onset of resistance, the two fluid model has been further used by Bennet et al. \cite{Bennet2012,Ullom_2015,Morgan2017} to make qualitatively good prediction of the logarithmic derivatives of
resistance with temperature and current ($\alpha$ and $\beta$) of TESs used in X-ray and gamma spectrometers.  
The original SBT or PSL model, which the two-fluid model is based upon, was later on used  by Bennet et al. to explain the bi-stable currents observed in some regions of the IV curve of TESs \cite{Bennett14}. Their finding might suggest
that, for TESs outside the weak-link regime, phase-slip
lines could be the mechanism for the observed resistance.

\subsection{Josephson Effects in DC and AC Biased TESs}
In the current and the following sections, we elaborate more on the implication of the Josephson effects on the performance of TESs.
The stronger evidence of the behaviour  of TES structures as superconducting weak-links  due to the long-range longitudinal proximity effect (LoPE) originating from the superconducting Nb leads is the peculiar dependence of the critical current  on the temperature $T$, $I_c(T)$, and on the magnetic field $B$ applied perpendicular to the direction of the current flow, $I_c(B)$. The former shows   the exponential dependence upon the TES length $L$ and the square root of the temperature difference $T-T_{ci}$:
\beq{WLIcT}
I_c(T,L)\propto \frac{L}{\xi(T)}e^{-\frac{L}{\xi(T)}},
\eeq
with $\xi(T)=\xi_i/(T/T_{ci}-1)^{1/2}$ for $T>T_{ci}$. 
 The latter, for $T>>T_c$, take a form similar to the well known Fraunhofer pattern characteristic of many Josephson structures \cite{Likharev79,BarPat1982}: 
\beq{WLIcB}
I_c(B)=I_{c0}\left|\frac{sin\left(\pi\frac{B}{B_0}\right)}{\pi\frac{B}{B_0}}\right|,
\eeq
The periodicity of the oscillations is $B_0=\Phi_0/(wL)$, where
$L$ is the junction length, $w$ the width and $\Phi_0 =2\pi\hbar/2e= 2.07\times 10^{-15}\,\mathrm{Wb}$  is the
flux quantum. The equation above assumes a sinusoidal current-phase
relationship, which corresponds to a uniform current density
distribution $J$ at zero applied field, and the presence of
negligible screening currents \cite{BarPat1982}.

The~longitudinal proximity effect was observed over extraordinary long ($L>100\,\mu \mathrm{m}$) distances and it is responsible for the enhancement, in~the proximity of the leads, of~the spatially varying superconducting order parameters $\Psi$ of the TES bilayer (\figref{fig:WLRSJ}a). It has been shown as well~\cite{Sadleir11} that the order parameter is suppressed in proximity to  normal metal structures deposited along or on top of the TES due to the lateral inverse proximity effect (LaiPE).

Both the effects described by \Eqsref{WLIcT}{WLIcB}, have been experimentally observed  over a broad range of TES sizes and geometries, both under DC and AC biasing \cite{Sadleir10,Sadleir11,Smith2014,Gottardi17,deWit_2020}.

The resistively shunted junction (RSJ) model, introduced above,   describes very successfully the macroscopic behaviour of  superconducting weak-links of many different kind. The RSJ model was first formalized for a dc-biased TES in the work of  Kozorezov~et~al.~\cite{Kozorezov11}, which shows how to derive the TES resistive transition analytically following the  Smoluchowski equation approach for quantum Brownian motion in a tilted periodic potential in the presence of thermal fluctuations~\cite{Coffey_2008}. The same approach can be used  to calculate exactly the  intrinsic TES reactance, which has important implications on the TES response, in~particular when operating under MHz voltage~biasing \cite{Gottardi_APL14}.

\begin{figure}
\center
\includegraphics[width=10cm]{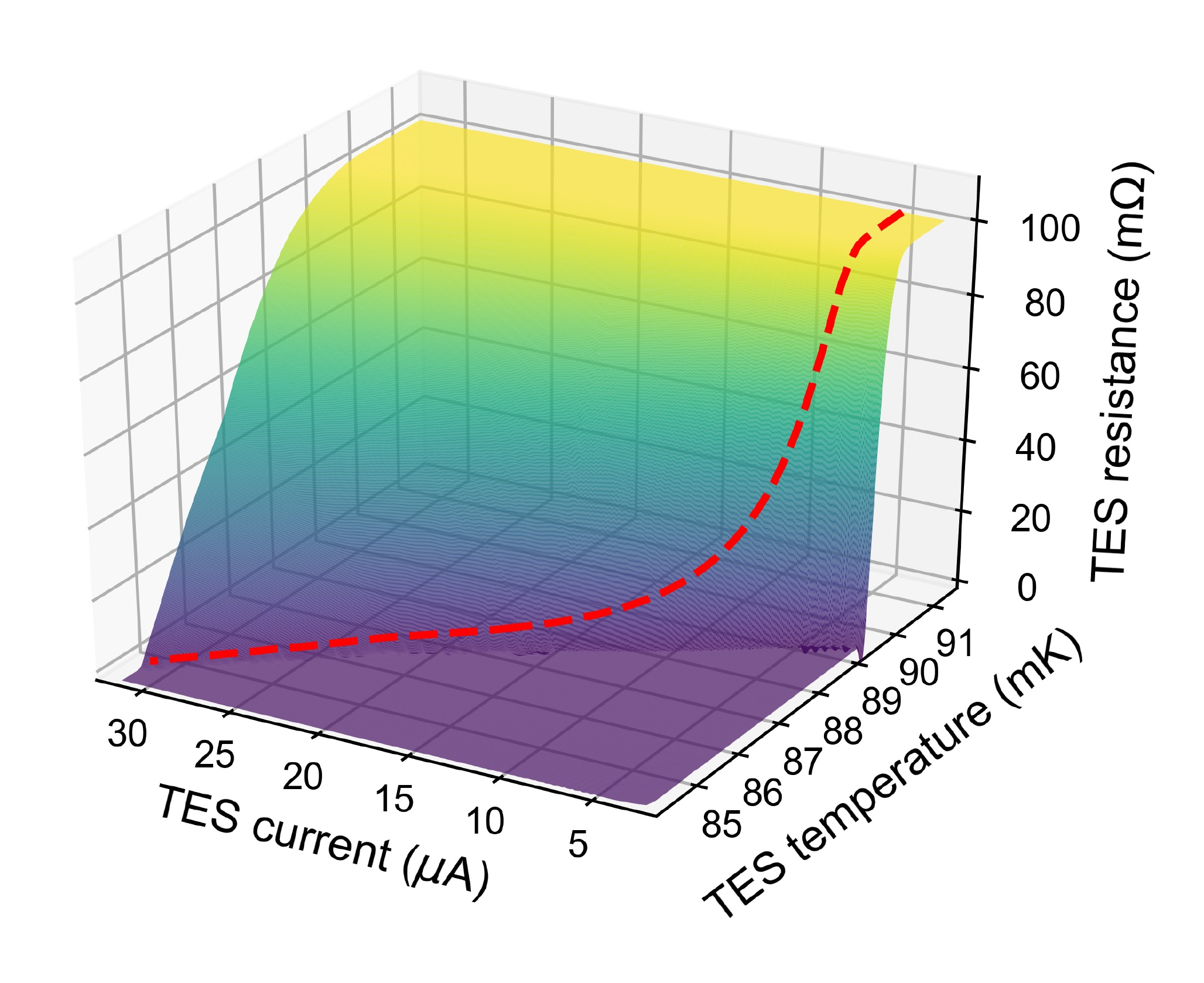}
\caption{\label{fig:RTI} Calculated resistive transition curve $R(T,I)$  for a TES microcalorimeter as described by Kozorezov et al.  \cite{Kozorezov11}. The dashed red line is the equilibrium curve across the transition following the detector power law.}
\end{figure}

The plot in \figref{fig:RTI} gives an~example of a TES resistive transition surface $R(T,I)$  calculated  following the approach described above. 

From the Josephson derivation in \Eqref{eq:RSJ}, the total current in the TES $I(t)$ is  the
sum of two components, the~Josephson current $I_{J}(t)$ and a quasi-particle (normal) current $I_{N} = V(t)/R_{N}$. 
In the most comprehensive case, the potential across the TES is the sum of a DC and AC voltage  $V_{TES}(t)=V_{dc}+V_{pk}\cos \omega_{0}t$. When the TES is dc biased,  $V_{dc}$ is  the constant voltage applied to the TES and the second term accounts for any potential ac excitation of peak amplitude $V_{pk}$ injected into the bias circuit. In~the  ac bias case, $V_{dc}=0$,  and~$V_{ac}=V_{pk}\cos \omega_{0}t$ is the applied voltage bias with amplitude $V_{ac}=V_{pk}$ and frequency  $\omega_{0} = 2\pi f_{bias}= \omega_{bias}$, that drives the TES into transition.  The~ac voltage across the TES forces the gauge invariant superconducting phase $\varphi$ to oscillate at $\omega_{0}$,  $\pi/2$ out-of-phase with respect to the voltage. The~$\varphi$ peak value depends on $V_{pk}/\omega_0$. 
\noindent From \Eqref{acJosephson}, the gauge invariant superconducting phase can be written as   $\varphi(t)=2e/\hbar\int V_{TES}(t)dt$ and  the Josephson current density becomes \cite{BarPat1982} 
\beq{Jjgeneral}
J_{J}(t)=J_c(T)\sin\left[\frac{2eV_{dc}}{\hbar}t+\frac{2eV_{pk}}{\hbar\omega_0}\sin \omega_0t+2\pi\frac{A_{eff}(B_{\perp,DC}+B_{\perp,AC}(t))}{\Phi_0}\right].
\eeq
where  $A_{eff}$ is the effective weak-link area crossed by  the dc ($B_{\perp,DC}$) and ac ($B_{\perp,AC}$) perpendicular magnetic field, respectively. The~$B_\perp$ field is generally a combination  of an external field and  the self-magnetic field generated by the current flowing in the TES and the TES leads \cite{Smith2014}.
The equation above is the generalized equation for the Josephson current density in a weak-link and shows the dependency of the current density on the DC and AC voltage across the weak-link and the magnetic field applied perpendicular to the current flow.
In the stationary case, $\frac{d\varphi}{dt}=0$, and in the presence of a perpendicular constant  magnetic field and uniform current distribution, \Eqref{Jjgeneral} reduces to \Eqref{WLIcB}, after  the integration over the weak-link area $A_{eff}=wL$ \cite{BarPat1982}. 

At zero  magnetic field and under strict DC biasing ($V_{dc}\neq 0,\, V_{ac}=0$), the phase $\varphi$ varies in time and an alternating current is generated, $J(t)=J_c(T)\sin\left[\frac{2eV_{dc}}{\hbar}t\right]$, with  $\omega_J= 2eV_{dc}/\hbar=2\pi\times 483.6\, [\mathrm{MHz}]\, V_{dc}/[\mu\mathrm{V}]$ equal to the Josephson oscillation frequency. 
In TES  micro-calorimeters, the typical dc bias voltage, $V_{dc}$, is of the order of 40 to 80 nV, depending on the TES geometry and resistance. For these voltages, the Josephson oscillations occur at  frequencies between 20 to 40 MHz, which are generally well outside the read-out bandwidth. 
We will see in the next section, however, that that mixed-down effects might occur, in particular in the noise, observable in the detector kHz bandwidth.

In the frequency division read-out scheme, the TES detectors are AC biased and $V_{dc}=0$ in \Eqref{Jjgeneral}. It can bee shown \cite{Gottardi_APL14,GottNaga2021}, that an alternating current is then generated through the TES $\pi/2$-out-off-phase with respect to the applied voltage $V_{ac}$. This is equivalent as  saying that a TES possesses an intrinsic reactance, which becomes non-negligible when exciting  the device at the MHz frequencies \cite{Gottardi_APL14,Kozorezov11}.

\begin{figure}
\center
\includegraphics[width=12cm]{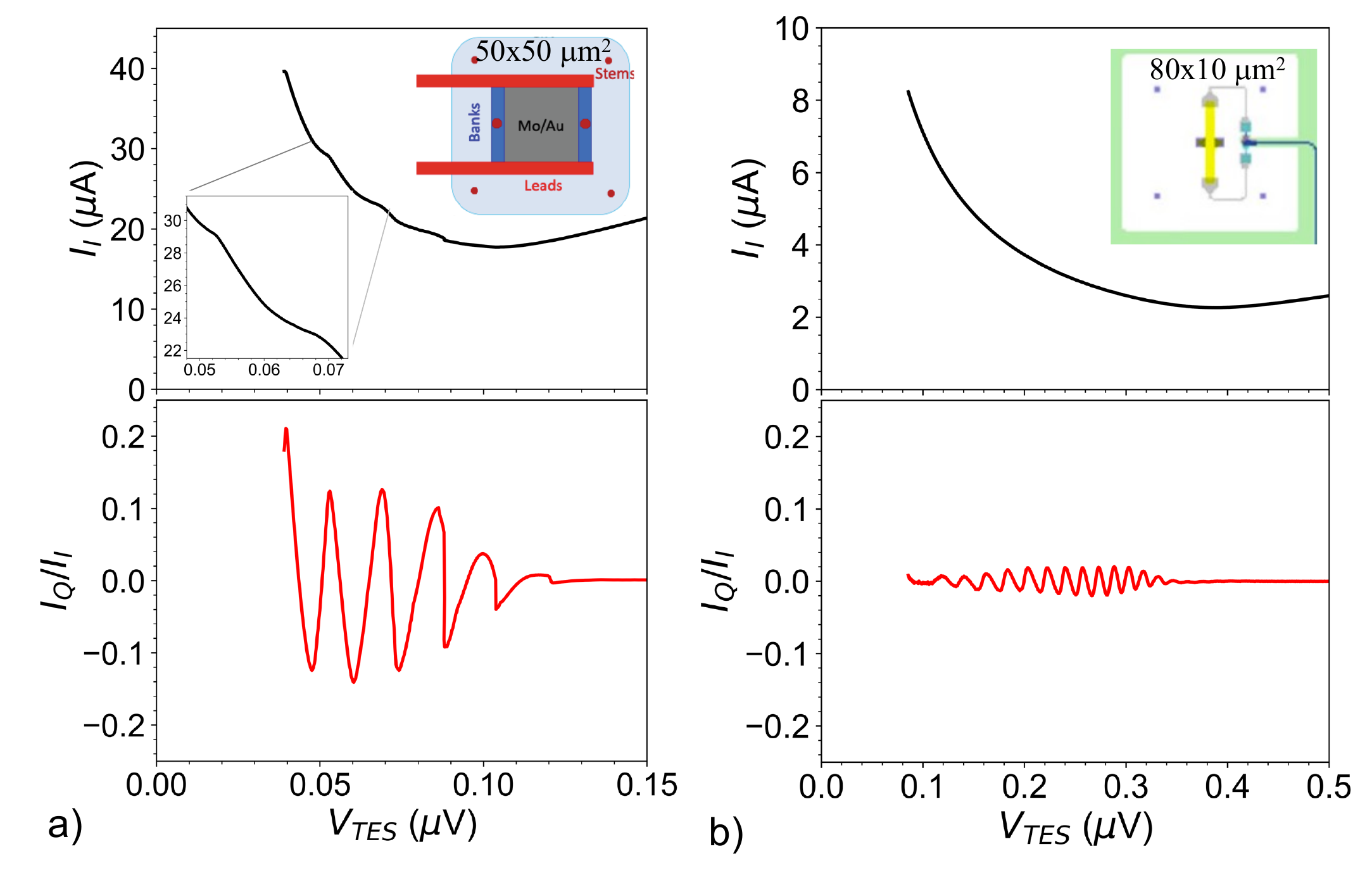}
\caption{\label{fig:acWL} a) Current-to-voltage characteristic of  a Mo/Au $50\times 50\, \mu\mathrm{m}^2$ (\mbox{AR = 1:1}, $R_N=8\,\mathrm{m}\Omega$) TES  X-ray calorimeter fabricated at NASA/GSFC. The pixel is AC biased at 4.0 MHz. The TES in-phase current, $I_I$, and the quadrature
current, $I_Q$ (Josephson current), normalized to $I_I$ are shown.  b) Mitigation of the weak-link effects for the AC bias read-out by employing the high normal resistance, high-aspect ratio devices developed at SRON. The graphs shows $I_I$ and $I_Q/I_I$ for a  Ti/Au $80\times 10\, \mu\mathrm{m}^2$ (AR = 8:1, $R_N\simeq 200\,\mathrm{m}\Omega$) TES X-ray calorimeter, AC biased at 4.0 MHz.}
\end{figure}

 In \figref{fig:acWL}.a,  the TES in-phase current $I_I$ and the quadrature
current $I_Q$ normalized to $I_I$ are shown as a function of the TES voltage for an AC-biased low resistance $50\times 50\, \mu\mathrm{m}^2$ MoAu TES  X-ray calorimeter fabricated at NASA/GSFC \cite{Miniussi2018}.  The quadrature component of the current is a direct measurement of the Josephson current described by \Eqref{Jjgeneral}. The steps observed in the in-phase current are the results of the intrinsic non-linear reactance expected when a TES behaves as weak-link \cite{Gottardi18}. 

As predicted by  the theoretical model presented in this session, the observed non-ideal behaviour of AC biased devices, caused by the  Josephson effects, can be minimized  by increasing the TES normal resistances, $R_N$, and by designing the TESs with higher aspect ratio (AR) \cite{Gottardi18,deWit_2020}. In~TESs with higher $R_N$, operating at the same power $P_J$, the~gauge invariant phase difference across the Josephson weak link is maximized since $\varphi \propto \sqrt{P_{J}R}/f_{bias}$ and the TES is less affected by the Josephson effects.
In~\figref{fig:acWL}.b, we give an example of the mitigation of the Josephson effects achievable by using high aspect ratio TESs. 
We show the current--to--voltage characteristics of an  SRON Ti/Au \mbox{$80\times 10\, \mu\mathrm{m}^2$} (AR = 8:1, $R_N\simeq 200\,\mathrm{m}\Omega$), measured under AC bias. The device has a $P_{J}\sim$ 1--2 $\mathrm{pW} $, compatible with the X-IFU requirement and is biased at $f_{bias}=4.0 \, \mathrm{MHz}$. The amplitude of the Josephson current is drastically reduced in the long, high resistance devices, in contrast to the detector of \figref{fig:acWL}.a. These novel TES design leads to a smoother resistive transition, even at high bias frequency.   

The strong non-linear nature of the Josephson current described in \Eqref{Jjgeneral} could challenge the calibration of the detector response as a function of energy both under AC and DC bias. In the former case, non-linearity arises from the bias-dependent detector reactance, in the latter case from the strong dependency on the  external magnetic field. The Josephson effects have to be minimized, for both the read-out schemes, by a proper detector design. The optimization process is complex and still a subject of active research \cite{deWit2022,Vaccaro2022,Smith2016}.  
\begin{figure}
\center
\includegraphics[width=11.5cm]{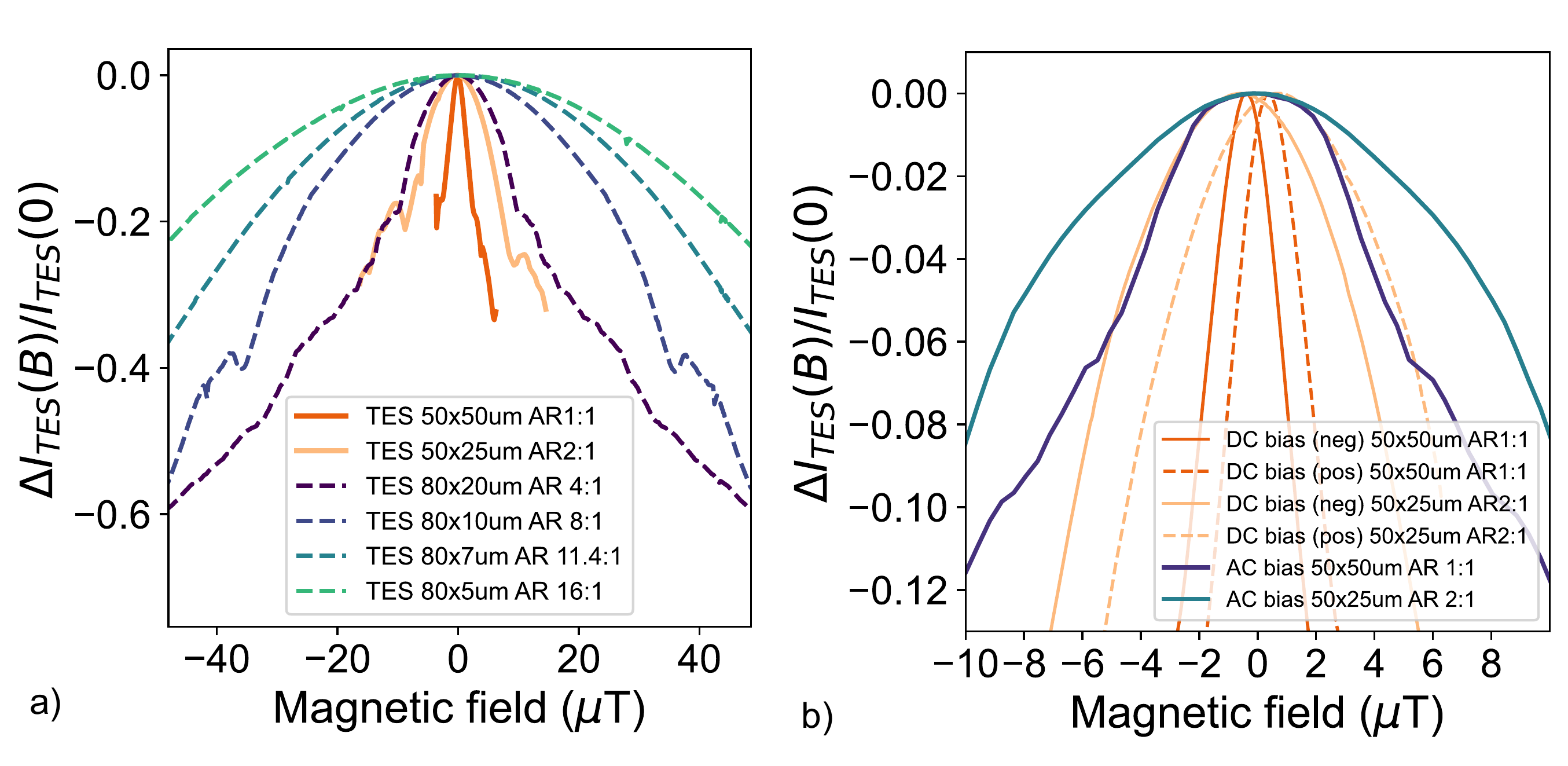}
\caption{\label{fig:IcB} a) Relative change in the TES  current as a function of the magnetic field perpendicular to the current flow for several SRON-fabricated Ti/Au TESs   with  low and high aspect-ratio geometry. The detectors are biased at their optimal value respectively under AC (bluish dashed lines) and DC voltage (orange lines). b) Magnetic field effect in pixels with low aspect ratio geometry operated respectively under AC (blue lines) and DC bias (orange lines). The solid and dashed lines of the DC bias case indicate respectively the negative and positive biasing.}
\end{figure}
As it is illustrated in \figref{fig:IcB}a), the high aspect ratio  design drastically reduces the dependency on the perpendicular magnetic field. The relative change in the TES current at the nominal bias condition is shown as a function of the magnetic field perpendicular to the current flow for SRON-fabricated devices with different aspect ratio.  The low aspect ratio  geometry (oranges lines) are typically used in the DC bias read-out due to their low normal resistance \cite{Smith2021}, while  the large normal resistance and high aspect ratio geometry (bluish dashed lines), envisioned  to minimized the detector reactance, have been used for the AC bias read-out \cite{Akamatsu2021}. 

The alternating current in the AC bias configuration mitigates the effect of the self-magnetic field observed in DC bias devices thoroughly discussed by Smith et al.\cite{Smith2014}. The effect is shown in \figref{fig:IcB}.b where identical pixels with low aspect ratio design have been measured both under AC and DC bias.

\subsection{Implication of the Weak-Link Behaviour on the Detector Noise}
\label{sec:RSJnoise}

The observed excess Johnson noise in TES-based detectors could have a natural explanation within the RSJ theory. Following the work of Likharev and Semenov~\cite{LikSem72} and Vystavkin {\it et al.}~\cite{Vyst74}, Kozorezov {\it et~al.}~\cite{Kozo12}  calculated the power spectral density of the voltage fluctuations, $S_V(\omega)$, across the TES (considered as a resistively shunted junction), averaged over the period of the Josephson oscillations. They obtained
\beq{eq:SvRSJ}
S_V(\omega)=\frac{4k_BT}{R_N}\sum_m|Z_{0m}|^2,
 \eeq
where $Z_{mm'}(\omega)$ are  the components of the impedance matrix of the biased TES with the index $m$ standing for the $m_{th}$ harmonic of the Josephson oscillation at $\omega_J$. This approach was used in Koch {\it et al.}~\cite{Koch80} to develop the quantum-noise theory of a resistively shunted Josephson junction and SQUID amplifiers. They showed that to~calculate  the total low frequency voltage fluctuations, one needs to take into account the  mixing down of high frequency noise  at harmonics of the Josephson frequency. 
Following these reasoning, in~\cite{Kozo12}, the authors argued that the only source of the intrinsic electrical noise in the TES is the  {\it equilibrium} Johnson normal current noise~\cite{LikSem72} enhanced by the nonlinear response of the weak link. Within this framework, there is no need to introduce the nonequilibrium Markovian fluctuation–dissipation relations as discussed in \cite{Irwin2006}.  The~RSJ model predicts a significantly higher broadband noise, with~respect to the non-equilibrium Johnson noise theory proposed by Irwin, at~the lower part of the resistive transition, as~generally observed in many experiments. 

A simpler expression  for \Eqref{eq:SvRSJ}, based on the approximations explained in~\cite{Kozo12}, has been recently derived by Wessel et al. \cite{Wessel2019,Wessels2021} in the form:
\beq{eq:SvRSJbeta}
\begin{split}
S_V(\omega)=4k_BTR\zeta_{RSJ}(I)\\
\mathrm{with}\;\;\; \zeta_{RSJ}(I)=1+\frac{5}{2}\beta+\frac{3}{2}\beta^2,
\end{split}
\eeq
where  $R_N$ in \Eqref{eq:SvRSJ} is replaced by $R$, given the fact that the thermal fluctuations are associated with the real part of the TES impedance at the equilibrium value. In~the same paper, the authors compare the measured Johnson noise for a few TES microcalorimeters~\cite{Wake2018} with  \Eqref{eq:SvRSJbeta}, the general form derived by Kogan and Nagaev~\cite{KogNag88} (KN) and  the prediction from the two-fluid model~\cite{Wessel2019,BennetPRB13}. In~the simplified form of the two-fluid model, no mixed-down noise from high  to low frequency is predicted, and~the expected noise is typically underestimated. A~better  agreement with the data is observed with the RSJ and the Kogan--Nagaev models. 

In~a recent study~\cite{Gottardi2021} on the noise of high-aspect ratio TESs under development at SRON for the MHz bias read-out, a~very good agreement between the observed Johnson noise and the prediction from the RSJ and Kogan--Nagaev has been demonstrated over a large number of TES designs and bias~conditions. The authors concluded 
that the fluctuation-dissipation theorem generalized for a nonlinear system in thermal equilibrium explains well the observed noise and that it is not necessary to introduce
the formalism for a nonlinear TES out of equilibrium described in \cite{Irwin2006}.
By assuming \Eqref{eq:SvRSJbeta} as the correct expression for the Johnson noise in a TES, de Wit et al. \cite{deWit2021} performed a study on  the impact of the internal thermal fluctuations driven by the TES-absorber coupling design. The work revealed subtle thermal effects and stressed the importance of a proper model for the TES Johnson noise.

\section{Detector Calibration Considerations} \label{Calibration}
The measured spectrum from an X-ray observatory will be a convolution of the astrophysical source with the response function of the instrument. In order to realize the exquisite spectroscopic capabilities of TES detectors on a real flight instrument, this response function must be properly calibrated and variations in the time due to changes in the instrument environment must be accounted for. Uncertainties in the knowledge of the overall system performance can lead to systematic errors in the data, and potentially compromise science objectives. Extensive calibration procedures, developed for Si thermistor microcalorimeters for the Astro-E and Astro-H missions \cite{Eckart2018}, can be applied to TES technology for missions such as Athena \cite{Cucchetti2018a,Cucchetti2018b}.

\subsection{Response Function}
The response function includes the effective area and line spread function (energy resolution and redistribution) as a function of energy. The effective area describes the X-ray collecting efficiency of telescope and includes contributions from the mirror, optical blocking filters and the detector. At the detector level the array quantum efficiency, including both the vertical stopping power of the absorbers and the geometric filling factor, must be precisely calibrated (at the few $\%$ level) as function of energy. This can be achieved through a combination of microscope inspection of the absorber size and gaps between pixels at various point across an array, measuring the change in mass of the detector wafer after the absorber deposition, and directly measuring the transmission of absorber parts at a synchrotron source.

\begin{SCfigure}
  \includegraphics[width=0.55\textwidth]{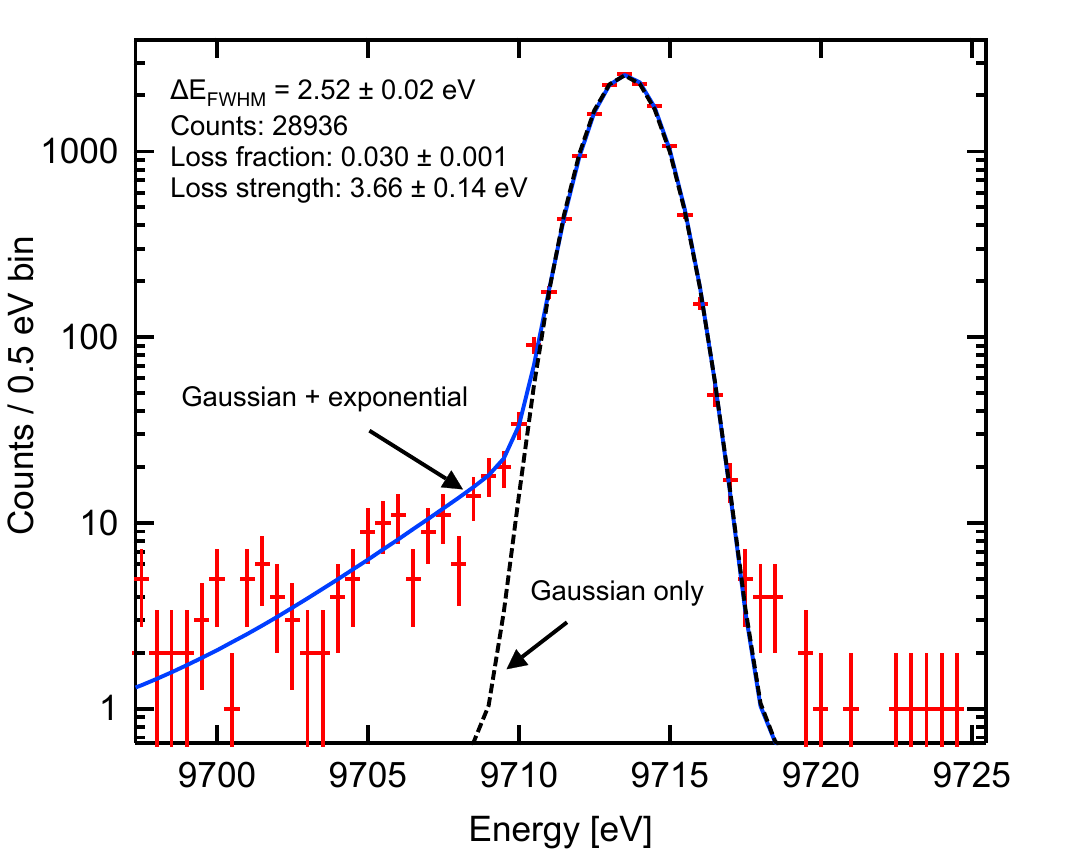}
  \caption{Example low energy tail below the main Gaussian core-LSF for 8 pixels in a prototype X-IFU array. The low energy tail accounts for 3$\%$ of events and has an exponential folding factor of 3.7 eV. These data was acquired using a channel cut crystal monochromator (CCCM) Au-L source \cite{Leutenegger2020} with intrinsic Gaussian line width of $<$ 0.5 eV.}
  \label{fig:core_LSF}
\end{SCfigure}

The line spread function (LSF) describes the probability that an incident X-ray event of energy $E$, will be measured in a given energy channel $E'$. The LSF consists of the Gaussian core and non-Gaussian or extended components. The core-LSF width is dominated by the intrinsic detector and system noise properties and is typically what is used to describe the energy resolution of the detector. Outside of the core, a small fraction of incident events appear re-distributed to lower energies due to various effects related to absorption physics. The re-distributed features or extended-LSF includes near core exponential tails in the spectrum just below the primary line, as well as electron-loss continuum and absorber escape peaks. Understanding and calibrating the LSF is important to be able to correctly interpret the measured spectrum from astrophysical objects. Uncertainties in the energy resolution will particularly affect measurements of the line broadening in astrophysical sources. Even with high purity electroplated Bi and Au absorbers, low level tails in the measured spectrum can typically be observed when using very narrow beam X-ray sources such as a channel cut crystal monochromator (CCCM) \cite{Leutenegger2020} or an Electron Beam Ion Trap (EBIT) \cite{ONeil2020}. An example spectrum is shown in Fig. \ref{fig:core_LSF} for a prototype Athena X-IFU detector which shows the Gaussian core and a residual low energy tail. Eckart et al. \cite{Eckart2019} studied the energy dependence of these residual low energy tails in similar detectors with Bi/Au absorbers and showed that the number of events redistributed to the low energy tail increased with decreasing photon energy.

The electron loss continuum occurs due to partial thermalization of events that are absorbed near the surface of the absorber. The initial photo-electron that is created by the primary X-ray may be scattered out of the absorber. Consequently only a small fraction of the incident X-ray energy is measured by the detector. This is empirically found to have approximately constant flux per energy interval below the primary X-ray line energy. Only a few $\%$ of the total events are redistributed to the electron-loss continuum.

\begin{figure}[h!]
\centering
  \includegraphics[width=1.0\textwidth]{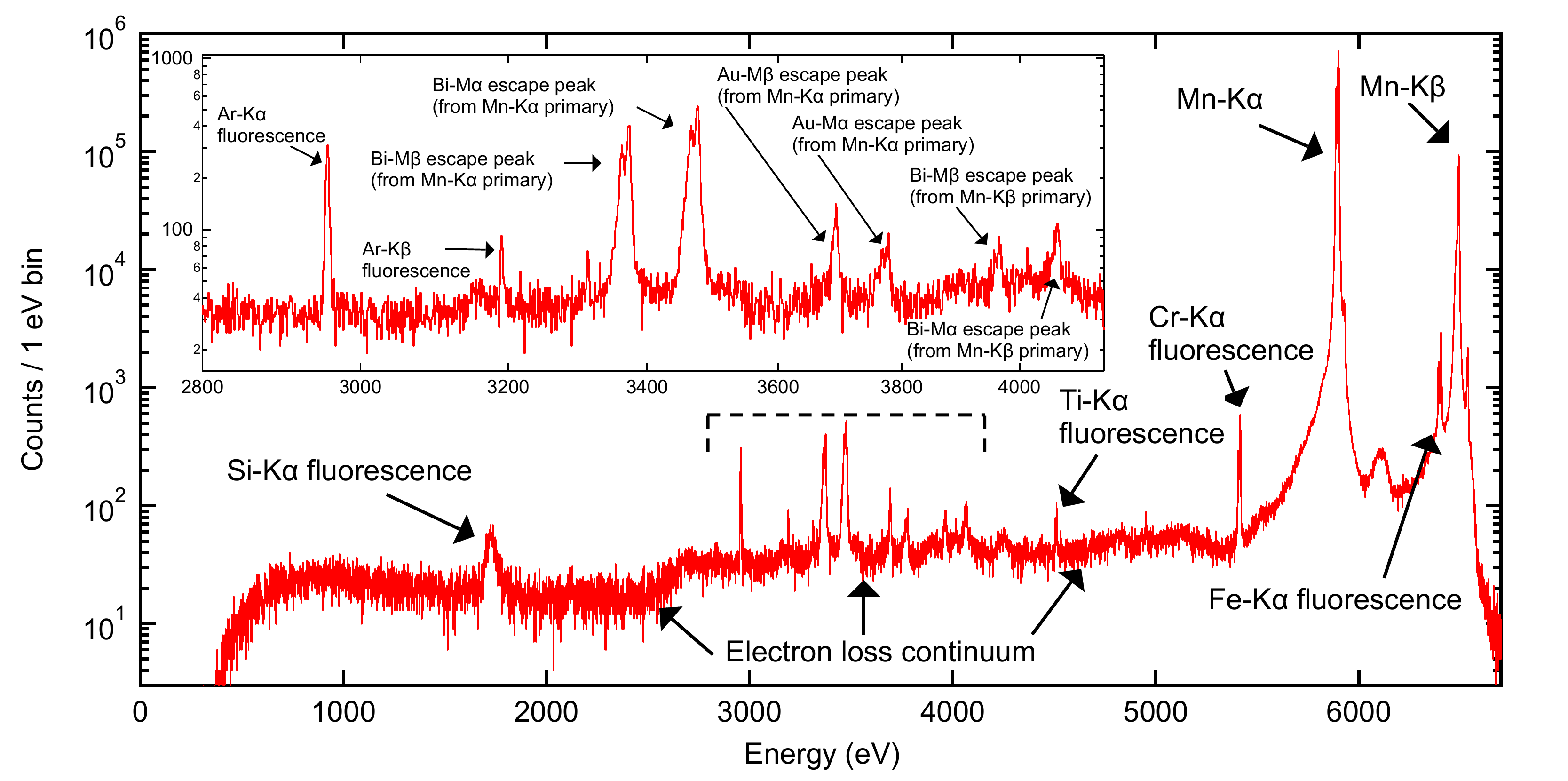}
  \caption{Mn-K spectrum from a prototype Athena/X-IFU array including over 200 pixels \cite{Smith2021}. The spectrum shows the primary Mn-K$\alpha$ and K$\beta$ lines from the X-ray source, and the background fluorescence and extended LSF features. This includes the electron loss continuum, fluorescence lines from Si (from detector wafer), Ar (from air), Ti, Cr and Fe (from the experimental apparatus), and Bi and Au escape peaks due to M-shell fluorescence. The inset shows a zoom-in of the region from 2.8-4.2 keV (indicated by the dashed line). The abrupt cut-off in the spectrum at 400 eV is due to the event trigger level and is not a property of the detector.}
  \label{fig:ELSF}
\end{figure}

When an X-ray event ejects an inner shell electron, the resulting vacancy may be filled by an outer shell electron and a fluorescent X-ray emitted. If this X-ray escapes the absorber, the detected energy is then equal to the primary X-ray energy less the fluorescent energy of the emitted X-ray. The resulting spectrum will contain a series of 'escape peaks' whose energies depend upon the primary X-ray energy and the absorber composition. In addition to these extended-LSF components, background peaks may be seen due to fluorescent X-ray emission from parts of the experimental apparatus exposed to the source flux. This could include the detector wafer, parts of the detector housing or cryostat and X-ray source itself. Figure \ref{fig:ELSF} shows a measured spectrum from a prototype Athena/X-IFU array \cite{Smith2021} including over 200 pixels. These devices have composite Bi/Au absorbers. The spectrum shows the primary Mn-K peaks from the source, as well as the extended-LSF and various fluorescent background peaks. Only the Si-K fluorescence line is fundamental to the detector. The main escape peaks are clearly visible above the electron loss continuum and are from Bi and Au M-shell fluorescent emission.

\subsection{Energy Scale and Sensitivity to Environmental Fluctuations}
\begin{SCfigure}
\centering
  \includegraphics[width=0.70\textwidth]{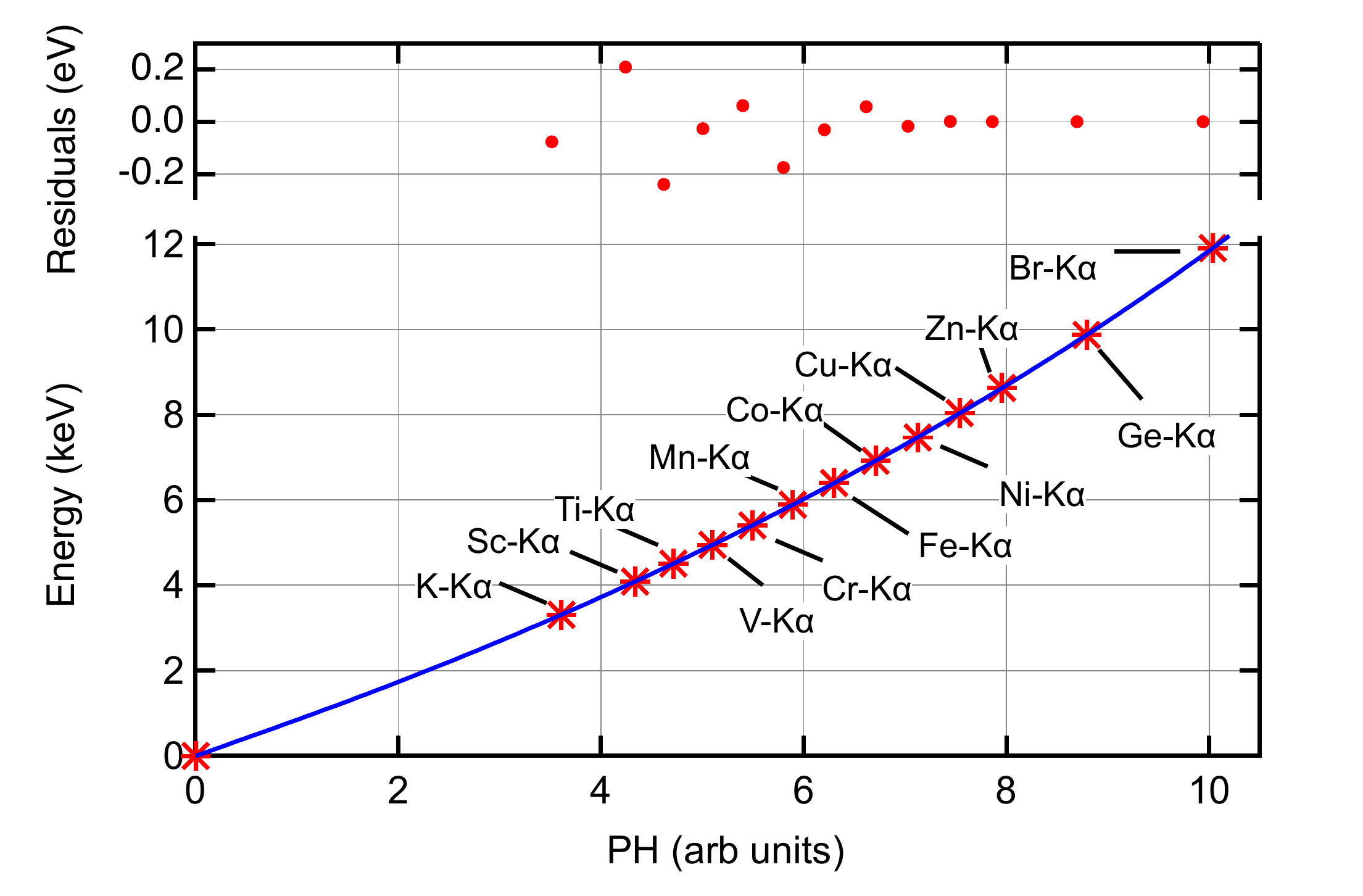}
  \caption{Example energy scale function $E(PH)$ for a prototype Athena/X-IFU pixel  measured over the range 3.3 - 12 keV using a rotating target source with 12 different fluorescent targets. Data is fitted with an $8^{th}$ order polynominal. For X-IFU, the energy scale must be known to a level of $\pm$0.4 eV for energies up to 7 keV.}
  \label{fig:energy_scale}
\end{SCfigure}

The energy scale is a function $E(PH)$ that relates the output of the event processor, $PH$ (the optimally filtered pulse height in engineering units), to calibrated energy units, $E$. TESs are inherently non-linear detectors and the superconducting-to-normal transition is very difficult to model accurately, thus an empirical approach to gain calibration is required. \figref{fig:energy_scale} shows an example of an energy scale function for a prototype TES detector for X-IFU. The gain function is found by fitting some function (such as a polynominal) to the known positions of the X-ray calibration points as function of optimally filtered pulse height. Gain errors can arise due to uncertainties in the line shapes and position of the calibration lines themselves as well as due to interpolation errors between the known calibration points.

Besides determining the gain function at the nominal detector operating points, it is important to understand how the gain evolves with the varying environmental conditions of the instrument. Variations in the detectors environment such as the heat-sink temperature, magnetic field, stray optical/IR power loading and temperature of bias/readout electronics  may result in drifts in the detector gain over time. If uncorrected this can result in resolution broadening and error in the absolute energy knowledge. Fluctuations in the TES voltage bias, $dV_b$, (due to a thermal drift in the electronics for example), will directly affect the bias point of the TES and therefore affect its gain. Fluctuations in the heat-sink temperature, $dT_{bath}$ will have a similar impact since it directly changes the operating point of the TES. But it can also slightly affect other properties such as the transition shape (due to the current dependence of the transition). Magnetic field, $B$, affects $R(T,I)$ via the temperature and field dependent critical current $I_C(T,B)$. Changes in $\alpha$ and $\beta$ directly affect the pulse shape and thus the gain of the detector. The sensitivity of a TES to $B$ depends on numerous factors including the details of the geometry, whether DC or AC biased, and the magnitude of any self-magnetic field \cite{Smith2014,Vaccaro2022,deWit2022}. For a device that shows weak-link behaviour (Fraunhoffer-like $I_C(B)$) the pulse-shape may oscillate with $B$ with the same periodicity seen in the $I_C(B)$ \cite{Smith2016,Smith13}. The effective area of the TES that threads magnetic flux ($A_{eff} = \Phi_0/B$) determines the periodicity in $I_C(B)$ and thus $PH(B)$. Thus a smaller TES has the potential to achieve less sensitivity to B-field. 

\begin{figure}[h!]
\centering
  \includegraphics[width=0.97\textwidth]{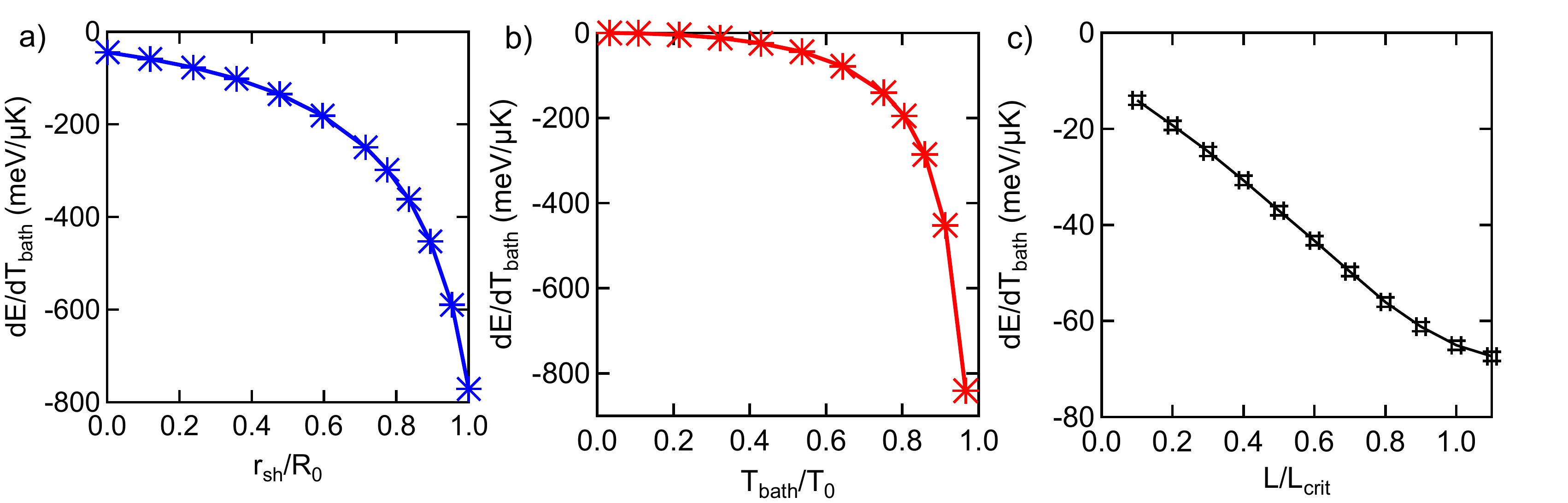}
  \caption{Simulated $dE/dT_{bath}$ as a function of a) $r_{sh}/R_0$, b) $T_{bath}/T_{0}$ and c) $L/L_{crit}$ for an X-IFU pixel at 6 keV. For Athena/X-IFU the baseline design is $r_{sh}/R_0 \sim$ 0.1 and $T_{bath}/T_0 \sim$ 0.6 and $L/L_{crit}$ $\sim$ 0.7.}
  \label{fig:dE_dTb}
\end{figure}

One of the most important properties of a TES that affects its gain sensitivity is the electrothermal feedback (ETF). The ETF that arises due to the voltage biased operation of the TES act as a restoring force, countering changes in external environment. Devices with high ETF, often parameterized by the 'loop gain' (see Section \ref{sec:ETF}), will generally have more immunity than devices operated in a low ETF regime. For a device with fixed transition properties, the only way to practically maximize the ETF is by ensuring the voltage bias shunt resistor is very much less than the TES operating resistance ($r_{sh}$ $\ll$ $R_0$) and/or the heat sink temperature is much less than the operating temperature of the TES ($T_{bath}$ $\ll$ $T_0$). As an example, \figref{fig:dE_dTb} shows the simulated $dE/dT_{bath}$ as a function of $r_{sh}/R_0$ and $T_{bath}/T_0$ for an Athena/X-IFU-like pixel. This illustrates how minimizing these ratios can be effective in reducing the sensitivity to bath temperature drifts. For a flight instrument, minimizing $r_{sh}$ has to be traded against other practical considerations such as the TES bias source range (lower $r_{sh}$ requires a larger current source to operate the TES at the same $R_0$), the power dissipation at the 50 mK stage (power dissipated in the shunt resistor scales as $R_0/r_{sh}$), and physical space constraints on the 50 mK stage (smaller $r_{sh}$ generally requires larger physical area). The lower the $T_{bath}$, the shorter the hold-time of the cooling chain adiabatic demagnetization refrigerator (ADR) becomes, unless the cooling capacity of the ADR is increased.

From the multiplexed read-out perspective, it is ideal to critically damp the pulses to minimize the requirements on the readout. The effect of adding inductance changes the pulse shape (increases the rise-time, reduces the fall-time and increases the pulse-height). However, this is non-linear with increasing inductance, with larger changes in responsivity occurring (for a given change in $L$) as $L/L_{crit}$ approaches unity. Consequently, the pulse shape also becomes more sensitive to changes in its environmental bias conditions. Thus, a side effect of larger inductance is an increased gain sensitivity (\figref{fig:dE_dTb} c).

\subsection{Drift Correction Algorithms}
On the final instrument, any drifts in gain will be monitored using fiducial line(s) generated by a calibration source \cite{deVries2017}. The measured drifts can then  be corrected during post-processing. The simplest method to correct the gain is to apply a linear stretch factor. This can work well when the detector gain is fairly linear and the scale of the drifts are relatively small ($\sim$ eV). However in the presence of larger drifts, and particularly for non-linear detectors such as TESs, large residual errors may arise at energies away from the fiducial energy. Porter et al. \cite{Porter2016} demonstrated a gain drift correction approach for microcalorimeter detectors (originally developed for the SXS instrument on Astro-H) that takes into account the non-linear shape of the energy gain scale to reduce gain errors over the full bandpass. In this method, reference gain scales are measured (during ground calibration) at three different heat sink temperatures, chosen to envelope the expected variation in operating conditions over the mission lifetime. Drifts are tracked at a single energy using the onboard calibration source and a new interpolated gain scale is derived (as a function of time) using the three reference calibration curves. Although the non-linear correction accounts for the shape of the gain scale, its evolution with variations in the environment may not be uniquely characterized using a single fiducial line. Using simulated TES data, Cucchetti et al. \cite{Cucchetti2018b} developed a multi-line, non-linear drift correction approach, which used two fiducial calibration lines (Cr-K$\alpha$ and Cu-K$\alpha$ for example) to track the gain instead of one which demonstrated reduced gain residuals across the bandpass. These multi-line approach is being developed for Athena/X-IFU. Cucchetti et al. \cite{Cucchetti2018c} also investigated a multi-parameter, non-linear drift correction approach that aims to incorporate  additional information from the detector baseline-level (the pre-trigger mean of the measured pulse) in the correction algorithm.




\section{Multi-Pixel TESs}
Several practical constraints  can limit the number of pixels in the array, both at the focal-plane assembly (FPA) and for the instrument as a whole. These include: the high density of wiring needed to route wires from the interior of the array to the bias circuit and readout components; the physical space required on the FPA for shunt resistors, Nyquist inductors and readout chips; the associated power dissipation of those additional components; the larger number of readout electronics boxes (which drives instrument mass and power), and the number of interconnects between temperature stages. Position-sensitive microcalorimeters are an alternative approach to arrays of individual TES pixels that enable a larger number of effective pixels in an array, without significantly increasing the complexity of the instrument. This comes with some trade in performance. Position-sensitive microcalorimeters have been proposed in various slightly different forms \cite{IYOMOTO2006,Stern2008, Mori2008, smith2004, Smith2009b}, but the basic principle is to use a single extended absorber, or series of absorbers, connected to one or more TES. Due to the finite thermal conductance between the position of the absorption event and the TES(s), a different characteristic pulse-shape is measured depending upon where the photon has been absorbed. It is this pulse-shape variation that enables position discrimination. This concept is often referred to as 'thermal multiplexing' which, when combined with an electrical multiplexed readout, is a very powerful tool for increasing the effective number of pixels in an array.
 
\begin{figure}[h!]
\centering
  \includegraphics[width=1.0\textwidth]{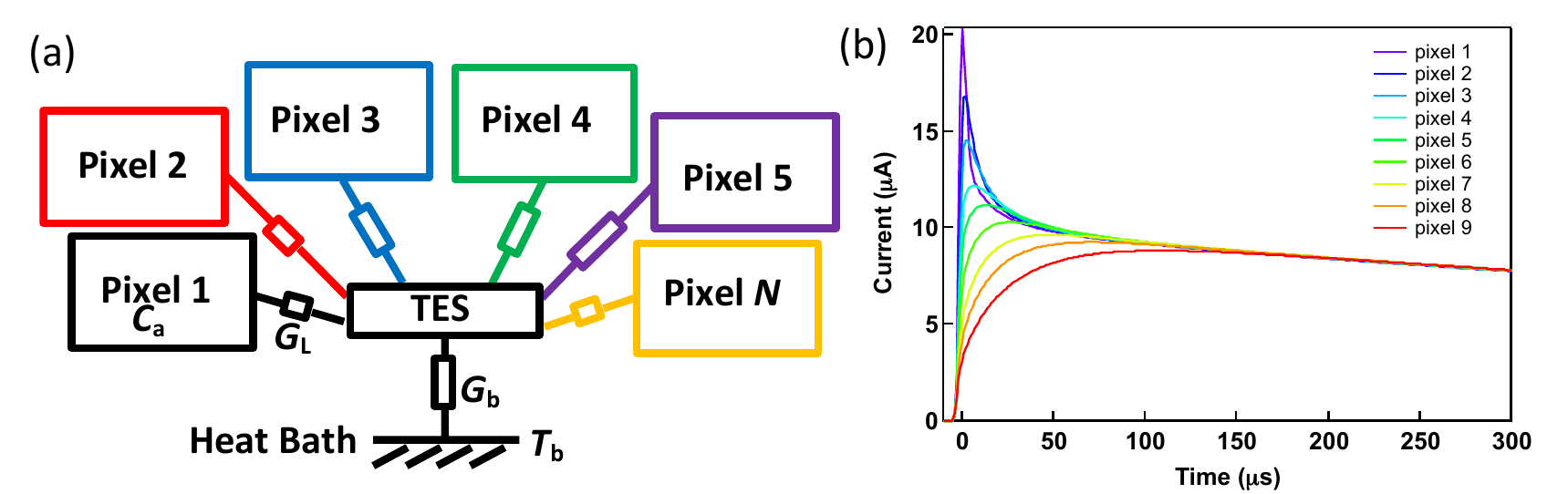}
  \caption{ a) Thermal design of an N-pixel hydra of the type under development at NASA GSFC. b) Measured average pulses from a 9-pixel hydra \cite{Smith2019}.}
  \label{fig:hydra1}
\end{figure}

Position-sensitive TESs for X-ray astrophysics were first developed at NASA/GSFC \cite{Figueroa2004, IYOMOTO2006}. The currently favoured design is referred to as a 'hydra'. The hydra consists of a single TES connected to series of discrete X-ray absorbers \cite{Smith2008}. Figure \ref{fig:hydra1} shows a thermal block diagram of the hydra concept consisting of N X-ray absorbers each with a different thermal 'link' to a TES. This configuration is referred to as a 'star' geometry. The links act as thermal low pass filters. By varying the thermal conductance of these thin metal links (by varying the length and width) the bandwidth of each thermal filter can be tuned to give a characteristic pulse shape for the signals from each of the X-ray absorbers. Decoupling pixels in this way enables position discrimination, but at some expense in energy resolution. The low-pass filtering of the pulse shape attenuates the signal relative to the detector noise and slightly degrades the energy resolution. The more thermally decoupled a pixel, the more degraded $\Delta E_{FWHM}$ is expected to be from the single-pixel limit. The decoupling of the absorbers also adds a small amount of internal thermal fluctuation noise (ITFN) between the absorber and the TES (as described in Section \ref{sec:noise}). If $\Delta E_{FWHM}$ becomes too degraded the ability to distinguish pulses will also be compromised. Thus the link conductances must be optimized to maximize the signal-to-noise while maintaining just enough separation between pulse shapes to determine position. The maximum count-rate capability of a hydra is also an important design consideration. The count-rate of a TES is typically limited by pulse pile-up measured in the TES data stream. Thus an N-pixel hydra will have the same maximum count-rate ability as a single pixel TES detector of the same physical size. However, the count-rate per imaging element will be N times lower. The larger the number of pixels in the hydra (the 'hydra factor'), the larger the performance degradation and the more technically complex the design is to implement. Hydras can be used for applications where it is desirable to achieve a much larger field-of-view, or improved angular resolution for the same field-of-view, when compared to an array of individual TES pixels (with some trade in performance). See Section \ref{missions} for proposed missions that include the use of hydras.

Figure \ref{fig:hydra1} also shows the measured average pulse shapes for a 9-pixel star-hydra developed at NASA/GSFC \cite{Smith2019}. The largest, fastest rising pulse shape corresponds to the pixel directly coupled to the TES. The pulse amplitude becomes smaller, and the rise-time slower, as each absorber is further decoupled from the sensor. After the initial, position-dependent equilibration signal, the hydra thermal networks comes in to equilibrium and the pulses from all pixels decay at the same exponential rate (determined by the electrothermal time constant of the hydra). These Mo/Au bilayer TESs had a transition temperature of 46 mK. Each Au X-ray absorber was 65 $\mu$m on the side and 5.2 $\mu$m thick. The achieved resolution was $\Delta E_{FWHM} = 2.23$ eV at Al-K$\alpha$, 2.44 eV at Mn-K$\alpha$, and 3.39 eV at Cu-K$\alpha$. Position discrimination was achieved from the rise-time of the pulses. 

To achieve a larger numbers of pixels / TES, it becomes increasingly more challenging to physically layout the internal links, whilst achieving the desired thermal conductances. As a solution, NASA/GSFC proposed a hierarchical 'tree' geometry. In the tree-hydra, instead of connecting every pixel directly to the TES, groups of pixels are connected together by link ‘branches’, and then 1 pixel from each group connects to the TES via link ‘trunks’. The more complex thermal network of the tree-hydras means that more sophisticated position-discrimination algorithms may be required to determine the event position. Prototype tree-hydras have been developed with up to 25 pixels per TES (5 trunks each with 5 branches) for NASA's Lynx mission concept \cite{Smith2020}. The authors were able to demonstrate position discrimination using a simple approach to parameterize the rising-edges with two metrics that extracts a fast and slow component to the rise-time. This allowed unique identification of every branch and trunk pixel. These tree-hydras demonstrated $\Delta E_{FWHM} = 1.66$ eV and $\Delta E_{FWHM} = 3.24$ eV for hydras with a 25 $\mu$m and 50 $\mu$m absorber pitch, respectively, for Al-K$\alpha$ X-rays \cite{Smith2020}. 

\section{Applications and Future Technology Needs}

\subsection{Ground Based Instrumentation} \label{Lab_astro}
X-ray spectroscopy contains diagnostic transitions from ions of every abundant cosmic metal and provides information on plasma temperature, density, elemental and ionic composition. The utility of these diagnostics requires knowledge of large sets of atomic data that describe the physics of the emission process. Ground based, laboratory astrophysics measurements are critical to provide the experimental knowledge needed to bench-mark the atomic models used to interpret high-resolution spectra from celestial observations. There are various facilities worldwide, which can generate laboratory plasmas in controlled conditions: this includes electron beam ion traps (EBIT) and tokamaks (used in fusion energy sciences). By varying the composition and ionization state of a hot gas generated under controlled conditions, one can relate observed X-ray spectra to the conditions of the gas, and then translate that knowledge to real astrophysical observations. The first EBIT to be developed was at Lawrence Livermore National Laboratory (LLNL) in 1985 \cite{Levine_1988}. EBITs have been extensively used to produce high quality atomic data including accurate wavelengths measurements, line intensity ratios, linear polarizations, and cross-sections to test and benchmark various models theories  \cite{Brown_2006,ONeil_2020,Gall_2020}. Several low temperature detectors systems have been deployed at the LLNL EBIT using silicon thermistor detectors developed from the Astro-E and Astro-H flight programs \cite{Porter2000,Porter2008}. A TES based detector system using prototype Athena X-IFU detector and readout technology will be deployed in the near future \cite{Smith2021}. A similar TES based spectrometer has been fielded at the NIST Gaithersburg EBIT \cite{Szypryt2019}.

In addition to the laboratory astrophysics measurements, X-ray TES microcalorimeters are now commonly employed at a variety of light sources, accelerator facilities, and laboratory-scale experiments carrying out measurements related to synchrotron-based absorption and emission spectroscopy and energy-resolved scattering; accelerator-based spectroscopy of hadronic atoms and particle-induced-emission spectroscopy; laboratory-based time-resolved absorption and emission spectroscopy; X-ray metrology. Further details of these instruments can be found in \cite{Ullom_2015,Doriese2017} and will not be discussed further here.
\subsection{Next Generation Space Mission Concepts}\label{missions}

In this section we briefly describe the TES detector concepts that are under development for the next generation of X-ray observatories as listed in Table \ref{tab:Xraymissions}. 
\begin{table}

\caption{\label{tab:Xraymissions} Overview of the characteristic of the TES-based instruments planned for the future X-ray space observatories. Athena is currently in the ESA Phase-A study. The~other missions are still in the selection or proposal~phase.}
\begin{tabularx}{1.\textwidth} { 
  | >{\centering\arraybackslash}X 
  | >{\centering\arraybackslash}X 
  | >{\centering\arraybackslash}X 
  | >{\centering\arraybackslash}X 
  | >{\centering\arraybackslash}X 
  | >{\centering\arraybackslash}X 
  | >{\centering\arraybackslash}X|}
 \hline

\textbf{\newline Mission}	& \textbf{ F.o.V.} &\textbf{Angular Resolution} &	\textbf{Number of Pixels} & \textbf{Energy} & \textbf{dE} & \textbf{Eff. Area \newline @ 1 keV}\\
\textbf{}&\textbf{(arc min)}& \textbf{(arc sec)}&& \textbf{(keV)} & \textbf{(eV)} &  (\boldmath$\mathrm{m}^2$)\\
\hline
Athena 		& 5& 5& $\sim$2,400			&0.2--12 &2--2.5 &1.4\\
LEM		&$30$&15& $\sim$16,000			&0.2--2&1--2		& $\sim$0.26\\
Lynx		&1--5&0.5--1& $\sim$100,000			&0.2--7 &0.3--3&0.2--2
\\
Super DIOS		&$>$30&0--15& $\sim$30,000	&0.2--2&$<2$	& $>$0.1\\
CWE		&$60$&$5$& $\sim$1 M			&0.1--3 &0.3&10\\

\hline
\end{tabularx}
\end{table}

Mature cryogenic X-ray microcalorimeter instruments using Si thermistor detectors have flown the XQC sounding rocket experiment and in orbit through the Japanese mission HITOMI, and will fly again in 2023 on XRISM. However, the higher performance/lower maturity TES based instruments are unlikely to be realized on a satellite mission until the 2030's. TES detectors have however been flown on the sub-orbital sounding rocket experiment Micro-X (Northwestern University, USA) in 2018 \cite{Adams2020}. Micro-X uses TES devices developed at NASA/GSFC with intrinsic energy resolution of 4.5 eV over the energy range of interest (0.1- 2.5 keV). The array consists of 128 pixels on a 600 $\mu$m pitch. The launch was intended to observe the supernova remnant Cassiopeia A. A failure in the attitude control system prevented the rocket from pointing and led to no time on target. However in-flight calibration data still provided important information about the performance of these detectors in a flight environment. The mission has recently (August 2022) had a  successful re-flight, observing Cassiopeia A, with results and analysis pending.

\begin{figure}[h!]
\centering
  \includegraphics[width=1.0\textwidth]{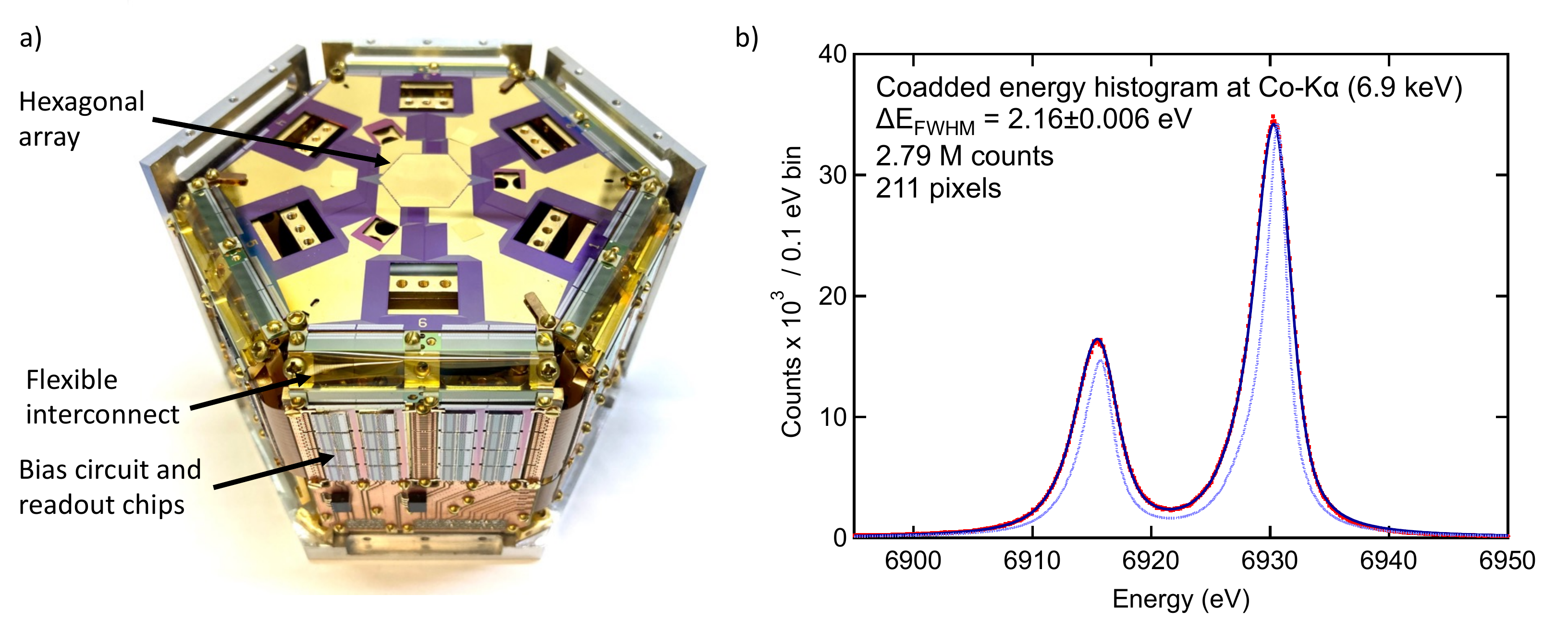}
  \caption{ a) Photograph of a full scale 3k-pixel array being developed at NASA/GSFC for Athena/X-IFU (image provided courtesy of NASA/GSFC) b) Energy histogram including 200-pixels readout with TDM from a 1k prototype array demonstrating $\Delta E_{FWHM} = 2.17$ eV for Co-K$\alpha$ X-rays (6.9 keV) \cite{Smith2021}. The blue-dashed line is the natural line shape of the Co-K$\alpha$ complex, and the solid blue-line is the fit to theh red data points.}
  \label{fig:XIFU}
\end{figure}

The Advanced Telescope for High-Energy Astrophysics (Athena) was selected as an L-class mission in ESA’s 2015-2025 Cosmic Vision plan and has an anticipated launch date in the late 2030's. Athena was chosen to address the science theme: ‘The Hot and Energetic Universe’, and as such will answer important questions related to the growth of black holes and large scale structure in the Universe. Athena is the most mature mission concept currently under development. As currently envisioned, Athena will have two main instruments, a wide-field imager (WFI) and an X-ray Integral Field Unit (X-IFU). The X-IFU \cite{XIFU2018} will comprise of a 2.4k TES pixel array (provided by NASA/GSFC) and will be read out with TDM (NIST/Boulder). The X-IFU will have a field-of-view of 5 arc-minutes and will provide imaging-spectroscopy over the energy range 0.2-12 keV, with an energy resolution of 2.5 eV at energies up to 7 keV. \figref{fig:XIFU} shows a prototype full scale X-IFU array with 3k pixels in a hexagonal configuration. The TESs are Mo/Au bilayers with a transition temperature of $\sim$ 90 mK fabricated on a 317 $\mu$m pitch. The X-ray stopping power is provided by Bi/Au absorbers $\sim$ 6 $\mu$m thick. Also shown in \figref{fig:XIFU} is an energy histogram showing the combined spectral performance at an energy of 7 keV and includes over 200 pixels from a prototype 1k array that was read out using TDM \cite{Smith2021}.

There are many design issues to consider when developing a complex instrument like X-IFU within the engineering limitations of a satellite. The packaging of the main sensor array chip, the cold read-out electronics stages, the long harness and the warm electronics  for a space-born instrument  \cite{Jackson2016,Geoffray2020,vanWeers2020} have several challenges related to the interaction between the different components and the thermal, mechanical, and electromagnetic environment. The focal plane assembly (FPA) has to be made robust against cosmic ray radiation \cite{Lotti2018,Miniussi_cosmicray2020,Peille2020,Hummatov2020}, infra-red and optical loading \cite{Barbera2018}, thermal fluctuation, magnetic field gradients, micro-vibration from the mechanical coolers \cite{Gottardi2019,vanWeers2020} and EMI from the satellite. The core of the FPA under development for X-IFU \cite{Jackson2016,Geoffray2020,vanWeers2020} consists of  the large TES microcalorimeters array and its SQUID-based multiplexing cold readout electronics, plus a second  TES array detector located at a close distance underneath the main array to be used as a cryogenic anti-coincidence detector \cite{Lotti2021,DAndrea2018}. The latter shall reduce the  particles background,  induced  by  primary  protons  of  both  solar  and  cosmic  rays  origin,  and~ secondary  electrons, to a total level  of $<$5$\cdot 10^{-3} \mathrm{cts}/\mathrm{cm}^{2}/\mathrm{s}/\mathrm{keV}$ in the 2--10 keV energy band. The other key building blocks of the FPA consist of the detector assembly, the  thermal suspension, the electromagnetic shielding, the  FPA-internal wiring harness, the  electronics assembly, and the thermal strapping. Critical to achieving high resolution with a TES X-ray spectrometer will be the shielding of the absolute static magnetic flux density component normal to the detector array surface, which needs to
be less than $1\,\mu \mathrm{T}$, including any gradients across the whole array. Additionally, the maximum normal magnetic field noise over the detector surface should
be less than  $20\, \mathrm{pT}/\sqrt{Hz}$ within the signal bandwidth. These stringent requirements call for a careful design of magnetic shields that should provide a shielding factor of $10^5$ or more \cite{Bergen2016}.

Orders of magnitude larger format pixel arrays are envisaged for the future proposed X-ray missions beyond Athena. To implement arrays greater than the k-pixel scale to 10's, 100's k-pixels or even mega pixel arrays will require further maturation of both focal plane array and readout technologies. In addition to hydras, other advanced focal plane technologies are being developed that could enable and expand the capabilities of next generation X-ray microcalorimeter instruments. This includes the development of advanced fabrication techniques that enable detectors of different designs to be fabricated in a single focal plane array, as well as high-yield, high density wiring. Hybrid arrays may include pixels of different pitch, absorber composition and TES transition temperature (or shape) and thus enable the flexibility to design different parts of the focal plane to achieve different performance requirements \cite{Wassell2017,Yoon2017}. This could enable more efficient instruments with enhanced performance capabilities. For example a point-source array optimized for high angular resolution, high count-rate observations could be accompanied by an array to expand the field-of-view for diffuse observations. The use of hydras can alleviate some of the issues associated with routing of high-density array wiring within the focal-plane array. However, extremely fine ($\leq$ 1 $\mu$m), high yield wiring is particularly important where small pixel pitches are desired or when single pixel TESs are preferred over hydras \cite{Datesman2017}. Multi-stack wiring layers buried directly within a Si wafer has been demonstrated by several commercial foundries such as Massachusetts Institute of Technology Lincoln Laboratory (MIT/LL) \cite{Tolpygo2015,Yohannes2015}. Multilayer technology may eventually enable other components like the bias (shunt resistors and inductors) and SQUID readout circuitry to be fabricated withing the same wafer as the detector. This may enable more compact focal-plane designs reducing the instrument mass (smaller focal-plane assemblies and magnetic shielding) or enable even larger arrays.

Readout technologies (see Chapter ”Signal readout for Transition-Edge Sensor X-ray imaging spectrometers”) are also rapidly advancing. Looking beyond current state of the art readouts such as time-domain-multiplexing (TDM) and low frequency-division-multiplexing (FDM), microwave (GHz) multiplexing ($\mu$-mux) is a rapidly maturing technology and is already fielded in ground based instruments \cite{Mates2017}. In $\mu$-mux each TES is coupled to an rf-SQUID and microwave resonator circuit. Each resonator is tuned to a different frequency and capacitively coupled to a single microwave feed-line. All pixels on the feed-line are then readout by a cryogenic amplifier such as a high-electron mobility transistor (HEMT). By moving the carrier signals to the GHz regime, $\mu$-mux dramatically increases the available bandwidth and allows many pixels to be read-out on a single microwave feed-line. HEMT amplifiers dissipate a significant amount of power and thus limit the number that can be implemented in a space instrument with finite available cooling power. Travelling-wave parametric amplifiers (TWPA) are promising alternatives to semiconductor amplifiers like HEMTs, with the potential to provide low noise amplification at 4K with significantly lower power dissipation \cite{Malnou2021}. Thus TWPA's have the potential to yet further increase the number of pixels that can be readout for future instruments.

\begin{figure}[h!]
\centering
  \includegraphics[width=0.95\textwidth]{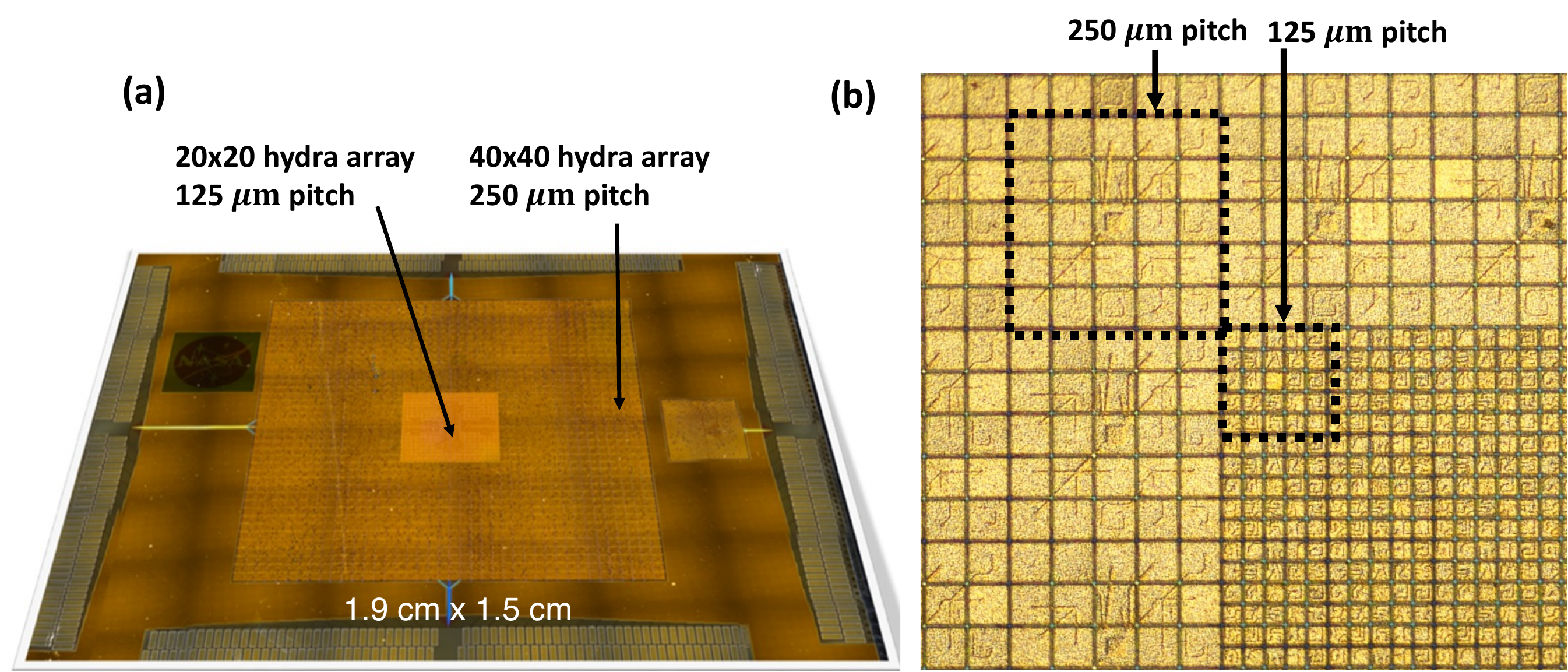}
  \caption{ a) Photograph of a prototype 50-k pixel Lynx Microcalorimeter Array under development at NASA/GSFC \cite{Bandler2019}. b) Zoom-in photograph of part of the LXM test array including 125 $\mu$m pitch Enhanced Main Array hydras in the central region of the array surrounded by a region of 250 $\mu$m pitch Main Array hydras \cite{Smith2020}.}
  \label{fig:lynx_array}
\end{figure}

The Lynx X-ray telescope was a Flagship mission concept studied by NASA as part of the 2020 Astrophysics Decadal survey \cite{Gaskin2019}. Although not selected to be the next Flagship, a Lynx-like mission was endorsed for further technology maturation and could launch sometime after 2040. The Lynx X-ray Microcalorimeter (LXM) is one of the main instruments that would combine a sub arc-second X-ray optic with a 100k-pixel array \cite{Bandler2019}. LXM is an example of a instrument that combines of many of the advance detector and readout concepts described above. LXM consist of a hybrid array of three different detectors on the same substrate. The LXM  main array consists of 3456, 25-pixel hydras on a 50 $\mu$m pitch (86k total pixels) with  $\Delta E_{FWHM} = 3$ eV at 0.2–7 keV. The main array provides a field-of-view of 5 arc-minutes and an angular resolution of 1 arc-second. The central 1 arc-minute region of the array consists 512, 25-pixel hydras with each pixel on a 25 $\mu$m pitch (13k total pixels) \cite{Smith2020}. This, enhanced main array will provide 0.5 arc-second imaging with $\Delta E_{FWHM} = 2$ eV up to 7 kev. Finally, there is an ultra-high resolution subarray of 25 $\mu$m single pixel TESs, optimized to achieve $\Delta E_{FWHM} = 0.3$ eV up to 1 keV \cite{Sakai2020}. NASA/GSFC has developed a process to incorporate planarized, multi-stack buried wiring layers from Massachusetts Institute of Technology Lincoln Laboratory (MIT/LL) with their TES array design. The LXM detector will be read-out with $\mu$-mux and HEMT amplifiers \cite{Bennett2019}. By combining 25-pixel hydras that incorporate buried wiring, with state-of-the-art $\mu$-mux, it becomes practical to realize a 100k-pixel instrument within the engineering constraints of a satellite. Figure \ref{fig:lynx_array} shows a prototype LXM array that includes all three array types.

The Line Emission Mapper (LEM) is a probe-class mission concept under study by NASA for a possible launch in the 2030's \cite{LEM2022}. LEM will combine a Si-shell mirror with an extremely large format microcalorimeter array to provide a $\sim$30 arc-minute field-of-view (FOV), 15 arc-second angular resolution, and $\sim$2 eV energy resolution. In contrast to ESA’s Athena and NASA’s Lynx concept, LEM will focus on large FOV science below 2 keV, such as the chemistry of the circumgalactic medium (CGM) and intergalactic medium (IGM). LEM's extremely large "grasp" (FOV x effective array) — enables the mapping of faint extended sources such as IGM, GCM, 12 times more efficiently than Athena. The LEM microcalorimter will consist of 4k, 4-pixel hydras (16k total imaging elements) and will be readout using TDM. A hybrid array configuration that includes the addition of a small sub-array of 1 eV resolution pixels is also under consideration.

The Super Diffuse Intergalactic Oxygen Surveyor (Super-DIOS) is mission concept under study in Japan (ISAS/JAXA) \cite{SuperDIOS2020} with a possible launch after 2030. The satellite would perform wide field X-ray spectroscopy with a comparable FoV of about 0.5–1 deg, and with a  angular resolution of $\sim$ 15 arc-seconds. The focal plane array will have 30k TESs read-out with $\mu$-mux, with resolution similar to X-IFU. The Cosmic Web Explorer (CWE) is a concept being proposed in Europe as part of ESA's Voyage-2050 program to survey the very faint warm-hot diffuse baryons in the local Universe \cite{Simionescu2019}. CWE will have a mega-pixel array with resolving power of 3000 at 1 keV, comparable to the LXM ultra high resolution array, but with larger collecting area and angular resolution.

\bibliographystyle{spbasic}
\bibliography{TESastrobook_main}

\end{document}